\documentclass[a4paper,11pt]{article}
\pdfoutput=1 

\usepackage{jcappub} 

\usepackage[T1]{fontenc} 
\usepackage{xspace}
\usepackage{multirow}
\usepackage{booktabs}
\usepackage{tabularx}
\usepackage{graphicx}
\usepackage{subcaption}
\usepackage{booktabs}   
\usepackage{array}      
\usepackage{makecell}   
\usepackage{cleveref}
\usepackage{comment}
\usepackage{rotating}
\usepackage[normalem]{ulem}

\newcolumntype{C}[1]{>{\centering\arraybackslash}m{#1}}

\usepackage{xcolor}
\definecolor{darkblue}{rgb}{0.0, 0.0, 0.55}
\definecolor{darkgreen}{rgb}{0.0, 0.55, 0.2}
\definecolor{darkred}{rgb}{0.55, 0.0, 0}

\newcommand{\lya}{Ly$\alpha$}
\newcommand{\lyaf}{Ly$\alpha$ forest}

\newcommand{\kmaxtwo}{\ensuremath{k_{\mathrm{max}} = 2\,h\,{\rm Mpc}^{-1}}}
\newcommand{\kmaxthree}{\ensuremath{k_{\mathrm{max}} = 3\,h\,{\rm Mpc}^{-1}}}

\newcommand{\Plya}{\ensuremath{P^{\alpha \alpha}}}
\newcommand{\Pcross}{\ensuremath{P^{\alpha \mathrm{m}}}}
\newcommand{\czeroF}{\ensuremath{c_0^{\alpha}}}
\newcommand{\ctwoF}{\ensuremath{c_2^{\alpha}}}
\newcommand{\cfourF}{\ensuremath{c_4^{\alpha}}}
\newcommand{\czeroDM}{\ensuremath{c_0^{\rm{m}}}}
\newcommand{\ctwoDM}{\ensuremath{c_2^{\rm{m}}}}
\newcommand{\cfourDM}{\ensuremath{c_4^{\rm{m}}}}

\title{Physical Calibration of a Minimal Effective Field Theory of the Three-Dimensional Lyman-$\alpha$ Forest}

\author[a,b,c,1]{Gabriele Autieri,\note{Corresponding author.}}
\author[d]{Vid Ir\v{s}i\v{c},}
\author[a,b,c,e]{Tomáš Šoltinský,}
\author[a,b,c,e]{Matteo Viel}

\affiliation[a]{SISSA - International School for Advanced Studies, Via Bonomea 265, 34136 Trieste, Italy}
\affiliation[b]{INFN – National Institute for Nuclear Physics, Via Valerio 2, I-34127 Trieste, Italy}
\affiliation[c]{IFPU, Institute for Fundamental Physics of the Universe, via Beirut 2, 34151 Trieste, Italy}
\affiliation[d]{Centre for Astrophysics Research, Department of Physics, Astronomy and Mathematics, University of Hertfordshire, Hatfield, AL109AB, UK}
\affiliation[e]{INAF - Osservatorio Astronomico di Trieste, Via G.B. Tiepolo, 11, I-34143 Trieste, Italy}

\emailAdd{gautieri@sissa.it}

\abstract{We study a minimal effective field theory description of the three-dimensional Lyman-$\alpha$ forest using the Sherwood and Sherwood--Relics hydrodynamical simulations. We model the Lyman-$\alpha$ flux auto-power spectrum and its cross-correlation power spectrum with the dark matter density field using a tree-level bias model supplemented by the leading counterterms and stochastic contributions. We find that the model describes the simulated auto- and cross-power spectra well up to $k_{\mathrm{max}}=3\,h\,{\rm Mpc}^{-1}$ and $k_{\mathrm{max}}=2\,h\,{\rm Mpc}^{-1}$, respectively. We analyse simulations spanning multiple redshifts, reionisation histories, box sizes and resolutions to assess the robustness of the model. Even within this minimal model, we find strong parameter degeneracies, highlighting the need for independent constraints on nuisance parameters in applications to real data analyses. The redshift evolution of the linear bias parameters is consistent with previous simulation-based studies and is driven primarily by the evolution of the effective optical depth, $\tau_{\mathrm{eff}}$. We also find that the inferred parameters are affected by the resolution of the simulation and, to a lesser extent, by the simulation box size. We also explore the impact of reionisation history, finding that variations mainly affect the linear bias parameters over the range of scales considered. Moreover, we find empirical correlations between model parameters and the \lya\ density bias that show some scatter, indicating that one single parameter is not enough to determine the model parameters. Finally, we compare our results with theoretical predictions from analytical models of the \lyaf. 
}

\begin{document}
\maketitle
\flushbottom

\section{Introduction}

The Lyman-$\alpha$ (\lya) forest is the dense series of absorption features observed blueward of the \lya\ emission line in the spectra of distant quasars, produced by neutral hydrogen in the intergalactic medium (IGM) along the line of sight \cite{Gunn:1965hd}. In the standard picture, most of the absorption arises from mildly overdense gas that traces the underlying matter distribution in a predictable way: photoionisation equilibrium with the ultraviolet background, together with the balance between photoheating and adiabatic cooling, establishes a tight relation between the gas temperature and density \cite{Hui:1997dp, Sanderbeck:2015bba}, so that the \lya\ optical depth is set by the local matter density through the so-called fluctuating Gunn-Peterson approximation \cite{Cen:1994da, Hernquist:1995uma, Bi:1996fh}. Because it probes the matter field on scales ($k\sim 0.1$--$100\,h\,{\rm Mpc}^{-1}$) and at redshifts ($z\sim 2$--$5$) that are difficult to access by other means, the \lyaf\ has become a key cosmological probe, complementary to other probes such as the cosmic microwave background and galaxy surveys \cite{Rauch:1998xn, Meiksin:2007rz}.
The clustering of the transmitted flux is sensitive to the amplitude and shape of the linear matter power spectrum, and the \lyaf\ has been used to constrain the primordial power spectrum \cite{Croft:1997jf, McDonald:1999dt, BOSS:2013rpr}, the sum of the neutrino masses \cite{SDSS:2004kqt, Palanque-Delabrouille:2014jca, Palanque-Delabrouille:2015pga, Yeche:2017upn, Palanque-Delabrouille:2019iyz}, the nature of dark matter (DM) \cite{Viel:2005qj, Viel:2013fqw, Dvorkin:2013cea, Irsic:2017yje, Armengaud:2017nkf, Baur:2017stq, Xu:2018efh, Garny:2018byk, Murgia:2018now,  Nori:2018pka,  Rogers:2020ltq, Villasenor:2022aiy, Irsic:2023equ, Garcia-Gallego:2025kiw}, and the thermal and reionisation history of the IGM \cite{Boera:2018vzq, Walther:2018pnn, Montero-Camacho:2019ucp, Molaro:2021tdz, Molaro:2023sys, Zheng:2025tjn}. Most of these analyses rely on the one-dimensional flux power spectrum, measured along individual lines of sight \cite{SDSS:2004kjl, eBOSS:2018qyj, Karacayli:2023afs}. This statistic is dominated by small, mildly non-linear scales, and its interpretation therefore requires comparison against hydrodynamical simulations of the IGM that resolve the low-density gas while spanning a sufficiently large volume \cite{Bolton:2016bfs, Puchwein:2022wvk, Walther:2024tcj}.

With the large quasar samples assembled by BOSS, eBOSS and, most recently, the Dark Energy Spectroscopic Instrument (DESI), it has become possible to measure correlations of the \lya\ absorption across neighbouring lines of sight, giving access to the full three-dimensional structure of the forest \cite{BOSS:2011ypt, Font-Ribera:2017txs, Abdul-Karim:2023sgf, deBelsunce:2024knf}. The three-dimensional flux power spectrum, $P^{\alpha \alpha}(k,\mu)$, where $\mu$ is the cosine of the angle between the wavevector and the line of sight, was first studied in detail in \cite{McDonald:2001fe}, where it was shown that on large scales it is well described by an anisotropic linear biasing relation analogous to the Kaiser formula for galaxies, controlled by a density bias $b_{\rm F}$ and a velocity-gradient bias $b_\eta$. The anisotropy of $P^{\alpha \alpha}(k,\mu)$ encodes redshift-space distortions and the Alcock-Paczynski effect, and it contains the baryon acoustic oscillation (BAO) feature. BAO in the \lyaf\ have provided some of the tightest constraints on the expansion history of the Universe at $z>2$ \cite{delubac2013baryon, BOSS:2013ola, BOSS:2014hwf, eBOSS:2020tmo}, recently culminating in a sub-percent measurement of the BAO scale by DESI \cite{DESI:2024lzq, DESI:2025zpo}. Beyond BAO, the broadband shape of the three-dimensional correlations carries additional cosmological information \cite{Cuceu:2021hlk, Cuceu:2022wbd}, and direct measurements of the 3D flux power spectrum from
data have now been performed \cite{deBelsunce:2024knf, Horowitz:2024nny, Karacayli:2025tlx}. Extracting this information requires an accurate model of $P^{\alpha \alpha}(k,\mu)$ down to mildly non-linear scales.

Two broad strategies exist for modelling the non-linear \lyaf\ flux power spectrum. The first calibrates a flexible fitting function against hydrodynamical simulations, multiplying the linear theory prediction by a correction that captures non-linear growth, peculiar velocities, thermal broadening and gas pressure \cite{McDonald:2001fe, Arinyo-i-Prats:2015vqa, Chabanier:2024knr}. While accurate within the simulated range, these corrections represent an empirical fitting function and are not derived from first principles. The second strategy is the effective field theory (EFT) of large-scale structure \cite{Baumann:2010tm, Carrasco:2012cv, Desjacques:2016bnm}, a systematic perturbative expansion in which the flux fluctuation is written in terms of all the operators allowed by the symmetries, with free coefficients (bias parameters, counterterms and stochastic terms) that encapsulate the effect of the unresolved small-scale physics. The expansion is controlled by the ratio $k/k_{\rm NL}$, where $k_{\rm NL}$ is the scale at which the density field becomes fully non-linear, and is therefore predictive only up to a limited wavenumber.

The EFT of the \lyaf\ was formulated at one-loop order in \cite{Ivanov:2023yla}, where it was shown that the bottom-up expansion of the transmitted flux coincides with the line-of-sight dependent bias model of \cite{Desjacques:2016bnm}, and that it reproduces the Sherwood simulations at the percent level up to $k\sim 3\,h\,{\rm Mpc}^{-1}$ at $z=2.8$. The same framework has since been applied to the eBOSS one-dimensional flux power spectrum to constrain the growth of structure and the neutrino mass \cite{Ivanov:2024jtl}, to the three-dimensional flux power spectrum measured from the ACCEL$^2$ simulations \cite{deBelsunce:2024rvv}, and extended to the cross-correlation of the \lyaf\ flux with halos \cite{Chudaykin:2025gsh} and to hybrid, simulation-based forward models
\cite{deBelsunce:2025bqc, deBelsunce:2025gci, deBelsunce:2026tks}. Its main appeal is that it provides a description rooted in symmetries, with a controlled regime of validity, at the cost of introducing a number of free parameters whose values and physical interpretation are not known a priori.

In this work we use the EFT framework to model the three-dimensional \lyaf\ flux power spectrum measured from the Sherwood and Sherwood--Relics hydrodynamical simulations \cite{Bolton:2016bfs, Puchwein:2022wvk}. In addition to the flux auto-power spectrum, $\Plya(k,\mu)$, we consider the cross-power spectrum between the \lyaf\ flux and the dark matter density field, $\Pcross(k,\mu)$, which shares the linear biases and counterterms of the auto-spectrum and helps to break degeneracies among the model parameters. Rather than adopting the full one-loop model, which carries a large number of free parameters, we explore a reduced description that retains the tree-level term together with the leading counterterms and stochastic contributions, and we assess which parameters are actually required to fit the data over the range of scales of interest. We then study how the inferred parameters depend on the simulation box size and resolution, how the linear biases evolve with redshift in comparison with previous simulation-based works, and how they respond to changes in the reionisation history, exploiting the Sherwood--Relics suite, which follows several models with different reionisation redshifts and different amounts of injected heat. We then examine empirical correlations between the EFT parameters and the linear density bias, which may help to reduce the number of free parameters in future analyses of the three-dimensional \lyaf, and finally we compare our results with theoretical predictions from analytical models of the forest.

The paper is organised as follows. In \autoref{sec:data} we describe the simulations and the variance-reduction technique. In \autoref{sec:model} we present the theoretical model, both the non-linear fitting-function approach and the EFT description of the flux auto- and cross-power spectra. We present our results in \autoref{sec:results} and we conclude and discuss some future prospects in \autoref{sec:conclusions}.

\section{Simulations}\label{sec:data}

In this work we use cosmological hydrodynamical simulations from the Sherwood\footnote{\url{https://www.nottingham.ac.uk/astronomy/sherwood/}}
\cite{Bolton:2016bfs} and Sherwood--Relics\footnote{\url{https://www.nottingham.ac.uk/astronomy/sherwood-Relics/index.php}} \cite{Puchwein:2022wvk} projects.
These were run with a modified version of the smoothed particle hydrodynamics code
\textsc{P-Gadget-3} \cite{Springel:2005nw} and were designed to follow the low-density IGM that gives rise to
the \lyaf, combining large volumes with the mass resolution needed to resolve the
small-scale structure of the gas. They adopt a flat $\Lambda$CDM cosmology with
$\Omega_{\rm m}=0.308$, $\Omega_{\rm b}=0.0482$, $h=0.678$, $\sigma_8=0.829$ and
$n_{\rm s}=0.961$. A detailed description of the simulations, and of the computation of
the \lyaf\ fields used here, is given in the companion paper, Šoltinský et al.\ (in
prep.). All distances and wavenumbers reported in the following are comoving, unless stated otherwise. We use Sherwood simulations with box sizes $L_{\rm box}=40$, $80$ and $160\,h^{-1}\,{\rm Mpc}$. For each box size, we consider simulations with $1024^3$ and $2048^3$ gas and dark matter particles, supplemented by a lower-resolution run with $L_\mathrm{box}=40\,h^{-1}\,{\rm Mpc}$ and $512^3$ particles. The $L_\mathrm{box} = 40\,h^{-1}\,{\rm Mpc}$ simulations with $512^3$ and $1024^3$ particles, together with the $L_\mathrm{box} = 80\,h^{-1}\,{\rm Mpc}$ simulation with $1024^3$ particles, are analysed at $z=2.4,2.8,3.2,3.6,4.2$ and $4.8$. All remaining runs are considered only at $z=2.4$.

The Sherwood--Relics suite extends the original Sherwood project with a non-equilibrium
thermo-chemistry solver and a set of models that vary the reionisation and thermal
history of the IGM \cite{Puchwein:2022wvk}. We use the models in which hydrogen
reionisation ends at $z_\mathrm{rei}^{\mathrm{end}}=5.37$, $6.0$, $6.7$ and $7.4$, labelled \textit{late}, \textit{reference}, \textit{early} and \textit{very early}, respectively. We also use two additional models, denoted \textit{hot} and \textit{cold}, in which the neutral hydrogen photoheating rate has been increased/decreased by a factor of $2$ compared to the reference, while keeping $z_\mathrm{rei}^{\mathrm{end}}=6.0$. We consider two box sizes, $L_{\rm box}=40$ and $80\,h^{-1}\,{\rm Mpc}$, each with $1024^3$ gas and dark matter particles. For the $40\,h^{-1}\,{\rm Mpc}$ box we analyse all six reionisation models at $z=2.4$ and $3.2$, while for the $80\,h^{-1}\,{\rm Mpc}$ box we analyse the \textit{reference}, \textit{hot} and \textit{cold} models at $z=2.4,\,2.8,\,3.2,\,3.6,\,4.2$ and $4.8$. We refer the reader to \cite{Bolton:2016bfs, Puchwein:2022wvk} for further details on the Sherwood and Sherwood--Relics simulation suites.

The runs used in this work are summarised in \autoref{tab:data_simulations}. In what follows we will refer to the simulations with the name specified in the table. 

\begin{table*}[tbp]
\centering
\caption{\label{tab:data_simulations}
Summary of the Sherwood and Sherwood--Relics simulations used in this work.
From left to right, the columns list the simulation name, the box size in $h^{-1}\,\mathrm{Mpc}$, the number of particles, the redshift of reionisation end (defined as the redshift when the volume-averaged neutral fraction in the simulation falls below $10^{-3}$), the gas temperature at the mean density, $T_0$, and the redshifts to which we have access, for which we define $z_\mathrm{all}=\{2.4, 2.8, 3.2, 3.6, 4.2, 4.8\}$. For more information on the simulation suites, we refer the reader to \cite{Bolton:2016bfs} and \cite{Puchwein:2022wvk}. 
}
\renewcommand{\arraystretch}{1.15}

\begin{tabular}{lccccc}
\toprule
Name
& $L_{\rm box}$
& $N_{\rm part}$
& $z_{\rm rei}^{\rm end}$
& $T_0(z=4.6)$
& Redshifts \\
&
$[h^{-1}\,{\rm Mpc}]$
&
&
& $[{\rm K}]$
& \\
\midrule

S40-512-ref
& 40
& $2\times 512^3$
& 15
& $8000$
& $z_{\rm all}$ \\

S40-1024-ref
& "
& $2\times1024^3$
& "
& "
& " \\

S40-2048-ref
& "
& $2\times 2048^3$
& "
& "
& 2.4 \\

S80-1024-ref
& 80
& $2\times1024^3$
& "
& "
& $z_{\rm all}$ \\

S80-2048-ref
& "
& $2\times2048^3$
& "
& "
& 2.4 \\

S160-1024-ref
& 160
& $2\times1024^3$
& "
& "
& " \\

S160-2048-ref
& "
& $2\times2048^3$
& "
& "
& " \\

\midrule

R40-1024-ref
& 40
& $2\times1024^3$
& 6.0
& $10\,066$
& $2.4,\ 3.2$ \\

R40-1024-hot
& "
& "
& "
& $13\,957$
& " \\

R40-1024-cold
& "
& "
& "
& $6598$
& " \\

R40-1024-late
& "
& "
& 5.37
& $10\,069$
& " \\

R40-1024-early
& "
& "
& 6.7
& $10\,050$
& " \\

R40-1024-very early
& "
& "
& 7.4
& $10\,003$
& "\\

R80-1024-ref
& 80
& $2\times 1024^3$
& 6.0
& $10\,066$
& $z_{\rm all}$ \\

R80-1024-hot
& "
& "
& "
& $13\,957$
& " \\

R80-1024-cold
& "
& "
& "
& $6598$
& " \\

\bottomrule
\end{tabular}

\end{table*}

\subsection{The \lyaf\ flux power spectrum}\label{sec:data_p3d}

From each simulation we draw a regular grid of sightlines and compute, along each of them,
the \lya\ optical depth, $\tau$, and the corresponding transmitted flux $F=e^{-\tau}$. At every redshift the mean flux is rescaled so that the effective optical depth
$\tau_{\rm eff}=-\ln\langle F\rangle$ matches the observed evolution of \cite{Viel:2013fqw}. We then build the flux contrast $\delta_{\rm F}=F/\bar{F}-1$ on a three-dimensional grid and measure its auto-power spectrum, $\Plya(k,\mu)$, together with the cross-power spectrum, $\Pcross(k,\mu)$, between $\delta_{\rm F}$ and the dark matter density field, to which we have access from the simulation snapshots, which we interpolate using the Cloud-in-Cell (CIC) algorithm. The technical details of the optical-depth computation are described in Šoltinský et al.\ (in prep.).

\subsection{Reducing sample variance}\label{sec:ZCV}

Because we work with individual simulations of limited volume, the measured power spectra
are affected by sample (cosmic) variance, which is largest on the large-scale modes that
are most sensitive to the linear bias parameters. To reduce it we use the Zel'dovich
control-variate (ZCV) technique \cite{Kokron:2022iok, DeRose:2022zfu}, recently applied to
the \lyaf\ in \cite{Hadzhiyska:2025cvk}. The method subtracts the noise from the
observable of interest by exploiting a cheaper, highly correlated field whose mean is
known analytically. As a control field we use the Zel'dovich approximation (ZA) matter
field, generated from the same initial conditions as each simulation with the
\textsc{N-GenIC}\footnote{\url{https://www.h-its.org/2014/11/05/ngenic-code/}} code. The de-noised power spectrum is
\begin{equation}
P^{\alpha \alpha}_\mathrm{ZCV} = P^{\alpha \alpha}_{\mathrm{sim}} - \beta \Bigl(P^{\mathrm{ZA}}_{\mathrm{sim}} - P^{\mathrm{ZA}}\Bigr),
\label{eq:ZCV}
\end{equation}
where $P^{\rm ZA}_{\rm sim}$ is the power spectrum measured from the control field and
$P^{\rm ZA}$ is its analytic, sample-variance-free mean, computed with the \texttt{ZeNBu}
code\footnote{\url{https://github.com/sfschen/ZeNBu}}. The coefficient $\beta$ is in
general arbitrary, and we set it to the value that minimises the variance of the ZCV power
spectrum,
\begin{equation}
    \beta \equiv \beta^\star = \frac{\mathrm{Cov}[P^{\alpha \alpha}_{\mathrm{sim}}, P^{\mathrm{ZA}}_{\mathrm{sim}}]}{\mathrm{Var}[P^{\mathrm{ZA}}_{\mathrm{sim}}]}.
    \label{eq:beta_equation}
\end{equation}
With this choice, the strong correlation between the simulated and ZA fields on large
scales is used to cancel the correlated part of the sample variance, substantially
reducing the uncertainties on large scales while leaving the signal unbiased. A more detailed description is given in Šoltinský et al.\ (in prep.); see also \cite{Hadzhiyska:2025cvk, Hadzhiyska:2026wts} for the control-variate technique applied to the \lyaf\ and to covariance estimation. 

\section{Modelling \lyaf\ non-linearities }\label{sec:model}

The observable we work with is the fluctuation of the transmitted flux fraction,
\begin{equation}
    \delta_{\rm{F}} = \frac{F(\mathbf{x})}{\bar{F}} -1, 
    \label{eq:deltaF}
\end{equation}
where $F(\mathbf{x}) = e^{-\tau(\mathbf{x})}$ is the transmitted flux, $\tau(\mathbf{x})$ is the redshift-space \lya\ optical depth and $\bar{F} = \langle F\rangle$ is the mean transmitted flux. The quantity of interest is the three-dimensional flux power spectrum, $\Plya(k,\mu)$, where $\mu$ is the cosine of the angle between the wavevector $\mathbf{k}$ and the line of sight.
The connection between the optical depth and the underlying matter field follows from the physical state of the absorbing gas. In the low-density IGM that produces the forest, the gas is highly photoionised and in ionisation equilibrium with the ultraviolet (UV) background. The neutral hydrogen abundance is therefore set by the balance between photoionisation and recombination. Since the hydrogen recombination coefficient scales approximately as $\alpha(T)\propto T^{-0.7}$, and neglecting peculiar velocities and thermal broadening, the real-space optical depth scales as \cite{Gunn:1965hd,Cen:1994da,Hernquist:1995uma,Bi:1996fh,Croft:1997jf}
\begin{equation}
    \tau(\mathbf{x})\propto n_{\mathrm{H}}^2 (\mathbf{x})\,T^{-0.7},
    \label{eq:temperature_scaling_of_tau}
\end{equation}
where the proportionality factor depends on the photoionisation rate, cosmology and the thermal and ionisation state of the gas. In practice, the overall normalisation is fixed by matching to the observed mean transmitted flux. On scales larger than the gas pressure smoothing scale, the baryonic density fluctuations trace those of the dark matter, thus the optical depth, and hence the transmitted flux, is a non-linear tracer of the underlying matter distribution. In redshift space, peculiar velocities displace the absorption along the line of sight, so that $\tau(\mathbf{x})$ in \eqref{eq:temperature_scaling_of_tau} has an additional dependence on the line-of-sight velocity field. 
However, this relation is only an approximation. It neglects thermal broadening, gas pressure and other processes that affect the transmitted flux. Rather than relying on this specific mapping, the flux fluctuation can be described through a general expansion in the matter density and velocity fields that, at linear order, reads
\begin{equation}
    \delta_{\rm{F}} = b_{\rm{F}}\delta + b_{\eta} \eta,
    \label{eq:deltaF_linear}
\end{equation}
where 
\begin{equation}
    \eta = -\frac{\partial_z v_z}{aH}
    \label{eq:eta_definition}
\end{equation}
is the normalised velocity gradient of matter along the line-of-sight. This is the most general linear-level expression since the only scalar quantities that can be constructed from the deformation tensor, $\phi_{,ij}$, which is equal to the second derivative of the gravitational potential, and the line-of-sight unit vector, $n_i$, are the trace of $\phi$, which is proportional to $\delta$, and $n_i n_j \phi_{, ij}$, which is proportional to $\eta$. Then, the biases $b_{\rm{F}}$ and $b_{\eta}$ are defined as the partial derivatives of $\delta_{\rm{F}}$ with respect to $\delta$ and $\eta$
\begin{equation}
    b_{\rm{F}} = \frac{\partial \delta_{\rm{F}}}{\partial \delta},\qquad b_{\eta} = \frac{\partial \delta_{\rm{F}}}{\partial \eta},
    \label{eq:biases_definition}
\end{equation}
where it is understood that partial derivatives are taken keeping the other variable fixed. Contrary to what happens for other biased tracers of the underlying matter field, denser regions and regions where the Hubble expansion rate has slowed down have stronger absorption and hence lower flux transmission, resulting in negative values for both $b_{\rm{F}}$ and $b_{\eta}$. By noting that in linear theory $\eta = f\mu^2 \delta$, where $f=d\ln D_+ / d\ln a$ is the linear growth rate and $D_+$ is the growth factor, we get the linear expression of the power spectrum of the \lyaf\ flux fluctuations 
\begin{equation}
    P^{\alpha \alpha}(k,\mu) = (K_1^{\alpha})^2 P_{\rm{lin}}(k),\quad K_1^\alpha \equiv (b_{\rm{F}} + b_{\eta} f\mu^2), 
    \label{eq:linear_flux_powerspectrum}
\end{equation}
where $P_{\rm{lin}}$ is the linear matter power spectrum and $K_1^\alpha$ is the linear kernel. Equation \eqref{eq:linear_flux_powerspectrum} has the same anisotropic, Kaiser-like form as for galaxies, with $b_{\rm F}$ and $\beta_{\rm F}\equiv f b_\eta/b_{\rm F}$ playing the roles of the linear bias and redshift-space distortion parameters. This expression holds only on large scales. To extend the description to smaller, non-linear scales, one usually multiplies the linear prediction by a correction that captures the non-linear physics of the IGM through a set of free parameters \cite{McDonald:2001fe, Arinyo-i-Prats:2015vqa, Chabanier:2024knr}.

\subsection{The effective field theory of \lyaf }
A more systematic approach is provided by the effective field theory (EFT) of large-scale structure \cite{Baumann:2010tm, Carrasco:2012cv, Desjacques:2016bnm}. Its main idea is that, on large scales, the relevant degrees of freedom are the long-wavelength matter density and velocity fields, while the complicated small-scale physics need not be modelled in detail. Their effect on large-scale statistics is captured by a finite set of free coefficients that multiply all the operators allowed by the symmetries of the problem, namely statistical homogeneity and isotropy and the equivalence principle, and are organised in a perturbative expansion controlled by the ratio $k/k_{\mathrm{NL}}$, where $k_{\mathrm{NL}}$ is the scale at which the matter density field becomes fully non-linear.
The \lyaf\ is a continuous field that is related to the underlying matter distribution through a non-linear transformation. For this reason the perturbative expansion contains, in addition to the usual operators built from the density and tidal fields, operators involving line-of-sight derivatives of the velocity, the leading one being the velocity gradient $\eta$ of \eqref{eq:eta_definition}. At lowest order the expansion reproduces the linear relation \eqref{eq:deltaF_linear}, while at higher orders further operators appear, each with its own free coefficient. This line-of-sight dependent bias expansion was formulated in \cite{Desjacques:2016bnm} and developed into an EFT of the \lyaf\ in \cite{Ivanov:2023yla}.
 
Interestingly, this construction can be reached in two equivalent ways. In the bottom-up approach one writes down, directly at the level of the flux, all operators compatible with the symmetries. In the top-down approach one starts from the exponential map $F = e^{-\tau}$, with $\tau$ being a function of the density and velocity fields and expands it perturbatively. In \cite{Ivanov:2023yla}, it was shown that, once the expansion is consistently renormalised, the two approaches give the same result. The renormalisation procedure introduces counterterms that absorb the sensitivity of the loop integrals to small scales, where perturbation theory breaks down. A further set of stochastic terms describes the small-scale fluctuations that are uncorrelated with the long-wavelength fields.
 
The EFT of the \lyaf\ has been tested against both simulations and data. It reproduces the Sherwood 1D and 3D flux power spectra at the sub-percent level up to $k\sim 3\,h\,{\rm Mpc}^{-1}$ at $z=2.8$ \cite{Ivanov:2023yla}, has been used to analyse the eBOSS 1D flux power spectrum \cite{Ivanov:2024jtl} and the 3D flux power spectrum of the ACCEL$^2$ simulations \cite{deBelsunce:2024rvv}, and has been extended to the cross-correlation of the \lyaf\ with halos \cite{Chudaykin:2025gsh} and to hybrid, simulation-based forward models \cite{deBelsunce:2025gci}. Because the expansion is controlled by $k/k_{\mathrm{NL}}$, its limit of validity is set by the non-linear scale, which can be estimated from the condition that the dimensionless linear power spectrum becomes of order unity,
\begin{equation}
    \Delta^2(k_{\mathrm{NL}}, z) = 1 \quad \Longrightarrow \quad k_{\mathrm{NL}} \simeq 3.6\, h\,{\rm Mpc}^{-1} \quad \text{for}\quad z=2.4.
    \label{eq:nonlinear_scale}
\end{equation}
For the \lyaf, two additional smoothing scales are relevant: gas pressure smoothing, set by the Jeans scale $k_J\sim 18\,h\,{\rm Mpc}^{-1}$ \cite{Gnedin:1997td, Garny:2020rom}, and thermal broadening of the absorption lines, with characteristic scale $k_S\sim 13\,h\,{\rm Mpc}^{-1}$ in the 1D power spectrum \cite{Garny:2020rom, Hui:1997dp, Theuns:1999mz}. At $z\sim 3$, both $k_J$ and $k_S$ are well above $k_{\mathrm{NL}}$, implying that the perturbative expansion breaks down before pressure smoothing or thermal broadening become important. Thus, the non-linear scale remains the limiting scale for the perturbative description. In the regime where $k_{\mathrm{NL}}$ sets the relevant perturbative cutoff, the one-loop EFT model for the \lyaf\ auto-power spectrum depends on $19$ free parameters: $2$ linear biases ($b_\mathrm{F}$ and $b_\eta$), $11$ non-linear bias parameters entering the EFT kernels, $3$ counterterms controlling the amplitude of $k^2 P_{\mathrm{lin}}$ corrections, and $3$ stochastic terms \cite{Ivanov:2023yla}. 
However, it is not yet clear whether all of these parameters are required to fit the data, nor at which scales the non-linear bias terms become relevant and should be included in the analysis. For this reason, we explore an alternative model for the \lyaf\ 3D flux power spectrum in which we truncate the perturbative expansion at tree level while retaining counterterms and stochastic terms. In this case, counterterms are expected to absorb the deviation from linear theory caused by omitting the one-loop contribution, therefore a significant contribution to these terms in the simulations would signal a possible need for higher-order terms in the expansion.
The model that we use for \Plya\ is 
\begin{equation}
\begin{aligned}
P^{\alpha \alpha}(k,\mu) ={}& (K_1^\alpha)^2 P_\mathrm{lin}(k)
-2(\czeroF+\ctwoF \mu^2 + \cfourF \mu^4)
K_1^\alpha \,k^2 P_{\mathrm{lin}}(k) \\
&+ P_{\mathrm{shot}}
+ s_0^\alpha \frac{k^2}{k_\mathrm{NL}^2}
+ s_2^\alpha \frac{k^2 \mu^2}{k_\mathrm{NL}^2}.
\end{aligned}
\label{eq:plya_model}
\end{equation}

A similar construction applies to the cross-correlation power spectrum of \lyaf\ flux and dark matter, with one important caveat. Matter and momentum conservation imply that the dark-matter stochastic terms start at $(k/k_{\mathrm{NL}})^2$, so no shot-noise term appears in the cross-spectrum model \cite{Mercolli:2013bsa, Senatore:2014eva}. We therefore model the cross-power spectrum as the symmetrised form of \eqref{eq:plya_model} under exchange of \lyaf\ and dark matter, excluding the shot-noise contribution:
\begin{equation}
\begin{aligned}
P^{\alpha \mathrm{m}}(k,\mu) ={}& K_1^\alpha K_1^\mathrm{m} P_\mathrm{lin}(k) 
-\Bigl[(\czeroF+\ctwoF \mu^2 + \cfourF \mu^4) K_1^{\mathrm{m}} \\
&+(\czeroDM+\ctwoDM \mu^2 + \cfourDM \mu^4) K_1^{\alpha}\Bigr]\,
k^2 P_{\mathrm{lin}}(k) \\
&+ s_0^{{\alpha \mathrm{m}}} \frac{k^2}{k_\mathrm{NL}^2}
+ s_2^{{\alpha \mathrm{m}}} \frac{k^2 \mu^2}{k_\mathrm{NL}^2},
\end{aligned}
\label{eq:pcross_model}
\end{equation}
 where $K_1^\mathrm{m} = (1 + f\mu^2)$ is the dark matter linear kernel, \czeroDM, \ctwoDM and \cfourDM are the dark matter counterterms and $s_0^{\alpha \mathrm{m}}$ and $s_2^{\alpha \mathrm{m}}$ are the cross-correlation stochastic terms. We note that, since the \lyaf\ is a non-linear transformation of the matter density, $\cfourF \neq \cfourDM$ \cite{Perko:2016puo, Chudaykin:2025gsh}, thus the cross-correlation power spectrum adds $5$ free parameters, the $3$ dark matter counterterms and the $2$ cross stochastic terms.

\section{Results}\label{sec:results}

In this work, we utilise the 3D auto-power spectrum of \lyaf, \Plya, and the cross-power spectrum of \lyaf\ and dark matter, \Pcross. The power spectra measurements as a function of $k$ and $\mu$ are obtained with the \texttt{Pylians3}\footnote{\href{https://github.com/franciscovillaescusa/Pylians3}{See \texttt{Pylians3} GitHub page.}} code and are then binned in $k$ and $\mu$. The $\mu$-space is divided into $4$ uniformly spaced bins in the interval $[0,1]$. Then, the ZCV method is applied to the data to reduce sample variance due to working with single simulations. To bin in $\mu$ the theoretical ZA power spectrum, we first compute the $\mu$-distribution of each $\mu$ bin for both spectra and use these distributions to randomly draw $N$ samples of $\mu$ values over which we average the theory prediction for each bin. We use $N=50$ as we find no improvement by increasing $N$ over that value. The same procedure is then followed to compute the theory predictions in the inference analysis. 
To sample the posterior distributions of the model parameters, we perform a Hamiltonian Monte Carlo (HMC) analysis. We make use of the No U-Turn Sampler (NUTS) algorithm \cite{Hoffman2011TheNS} for HMC implemented in the \texttt{NumPyro} library \cite{Bingham2018PyroDU, Phan2019ComposableEF}. We adopt a Gaussian likelihood, defined as
\begin{equation}
    -2\mathcal{L}_\theta = \Delta \Theta^{\mathrm T} \cdot \mathcal{C}^{-1} \cdot \Delta \Theta, 
    \label{eq:likelihood}
\end{equation}
where $\Theta$ is the observable vector, $\Delta \Theta \equiv \Theta^{\mathrm{theory}} - \Theta^{\mathrm{data}}$ is the difference between the theory and the data vector, and $\mathcal{C}$ is the covariance matrix. We compute the covariance matrix with the jackknife method as well as in the Gaussian approximation and, since we find no significant difference between the two in the parameter inference, in the following we will use the Gaussian covariance matrix. In the most general setup, the observables vector $\Theta$ is composed of the two different spectra $\Theta = (\Plya, \Pcross)$ and the covariance matrix is 
\begin{equation}
    \mathcal{C} =\left( \begin{array}{lr} C^{\alpha\alpha, \alpha\alpha} & C^{\alpha\alpha, \alpha \mathrm{m}}\\
    C^{\alpha \mathrm{m}, \alpha\alpha} & C^{\alpha \mathrm{m},\alpha \mathrm{m}}
    \end{array}
    \right),
    \label{eq:covariance_matrix}
\end{equation}
where each block represents a diagonal matrix, meaning that there is no covariance between different $k-$ and $\mu-$ bins. For a given $k-$ and $\mu-$ bin indexed by $i$, the blocks are defined as 
\begin{align}
    &C_{ii}^{\alpha\alpha, \alpha\alpha} = 2N_i^{-1} P^{\alpha \alpha}_i P^{\alpha \alpha}_i\\
    &C_{ii}^{\alpha \mathrm{m},\alpha \mathrm{m}} = N_i^{-1} (P^{\alpha \mathrm{m}}_i P^{\alpha \mathrm{m}}_i + P^{\alpha \alpha}_i P^{\mathrm{m}\mathrm{m}}_i)\\
    &C_{ii}^{\alpha\alpha, \alpha \mathrm{m}} = 2N_i^{-1} P^{\alpha \alpha}_i P^{\alpha \mathrm{m}}_i.
    \label{eq:covariance_matrix_blocks}
\end{align}
In the above expression, $N_i$ is the number of independent modes in each bin. In the inference, we choose wide uniform priors for all model parameters and, unless specified otherwise, we fit \Plya\ up to \kmaxthree\ and \Pcross\ up to \kmaxtwo.

\subsection{Fitting auto and cross consistently}
\begin{figure}[tbp]
\centering 
\includegraphics[width=\textwidth]{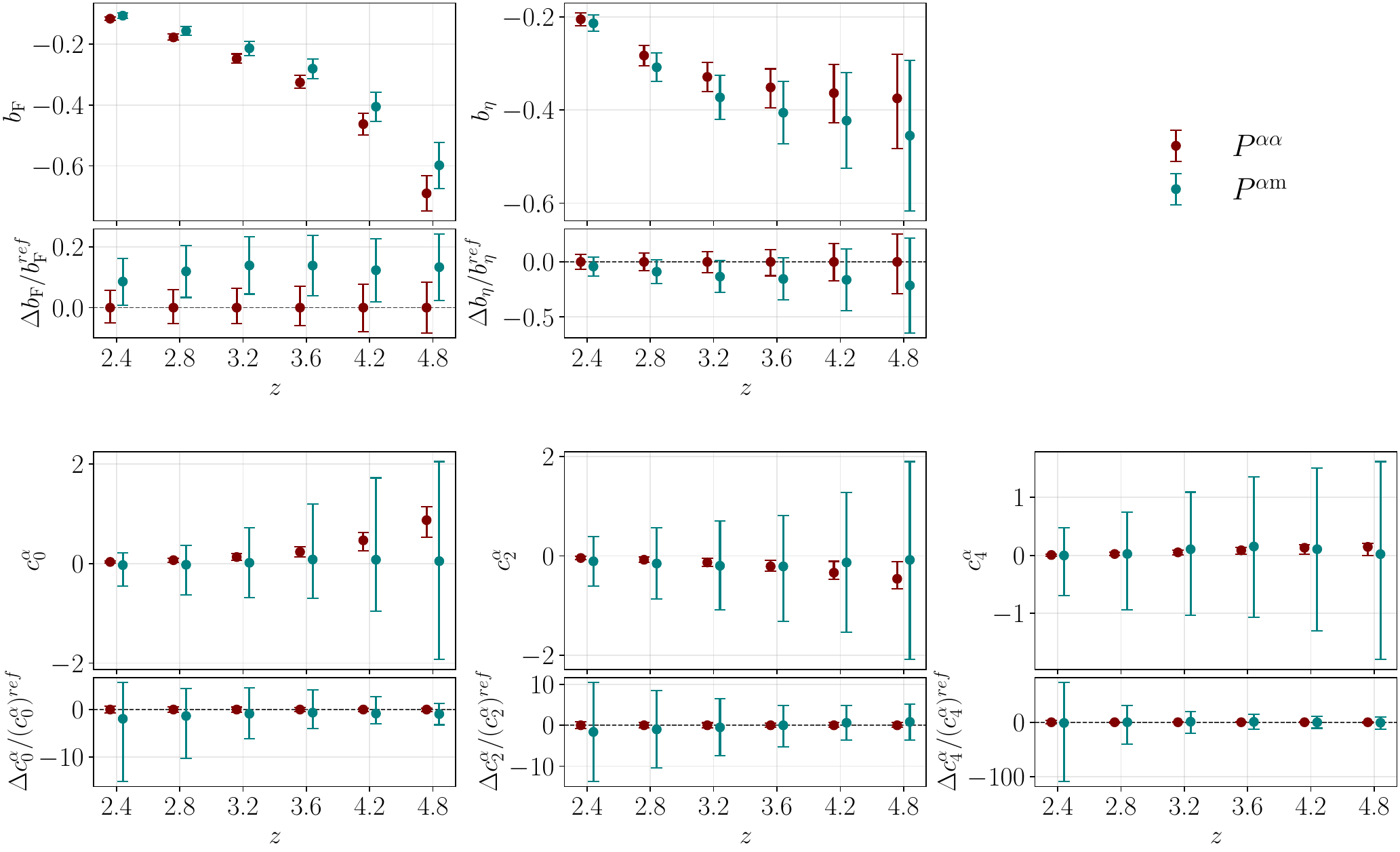}
\hfill
\caption{\label{fig:bestfits_auto_alone_vs_cross_alone}
Comparison of best-fit results for $b_\mathrm{F}$, $b_\eta$, $\czeroF$, $\ctwoF$ and $\cfourF$ obtained by fitting the \lyaf\ auto-power spectrum, $\Plya$, alone compared to fitting the \lyaf--DM cross-power spectrum, $\Pcross$, alone.}
\end{figure}

\begin{figure}[tbp]
\centering 
\includegraphics[width=0.8\textwidth]{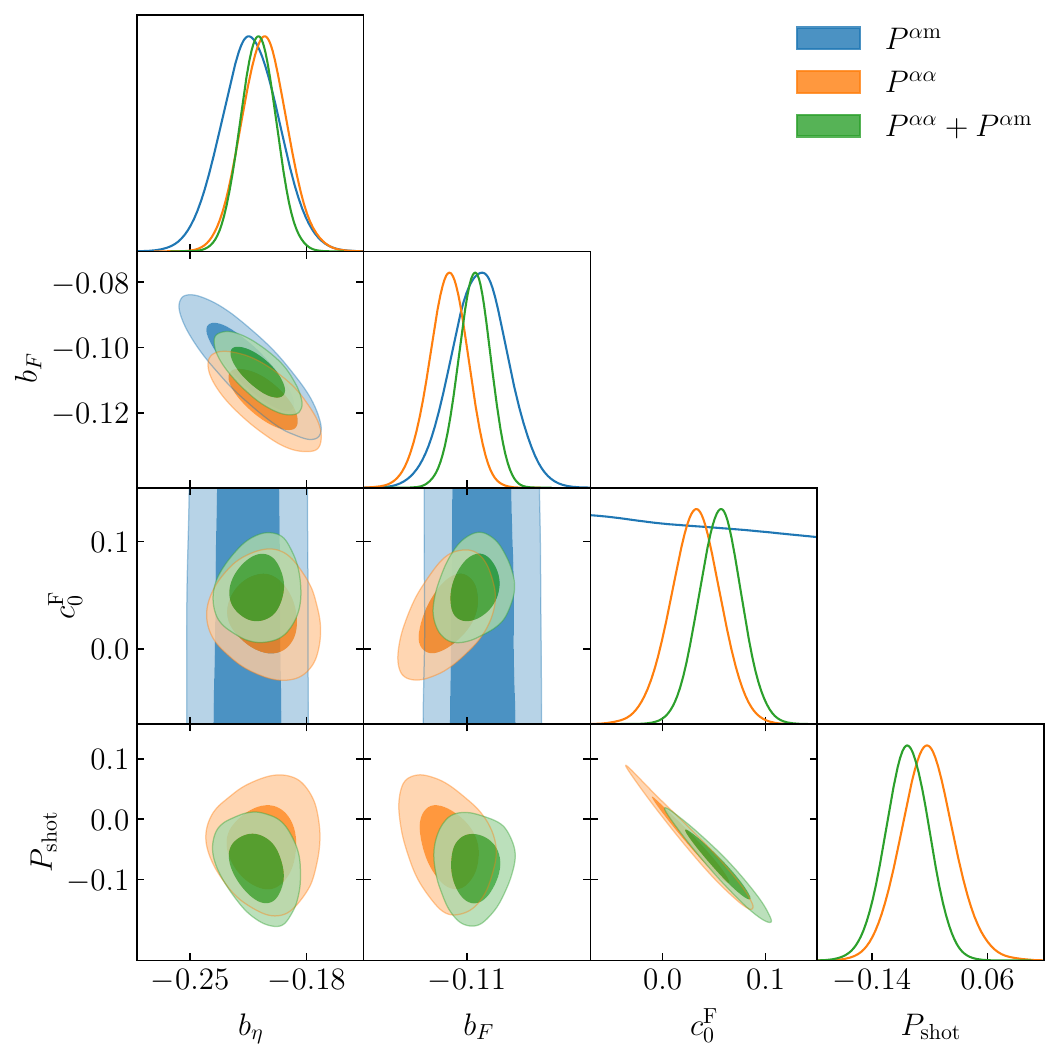}
\hfill
\caption{\label{fig:triangle_plot_auto_alone_vs_cross_alone_vs_combined}
Triangle plot for fitting \Plya\ alone, \Pcross\ alone and the two combined together. We plot here the results from the S40-1024-ref simulation at $z=2.4$.}
\end{figure}

We begin by discussing the consistency of fitting the \lyaf\ auto-power spectrum, \Plya, and the \lyaf--DM cross-correlation power spectrum, \Pcross, together. This discussion is necessary because it is well-known that one-loop corrections within the EFT framework are required to correctly capture non-linear effects in the power spectrum of the DM field. Additionally, it has been shown that non-linear peculiar velocity effects can be significant in halo--\lyaf\ correlations \cite{Givans:2022qgb}. However, in our modelling we do not include one-loop corrections in the cross-power spectrum. To mitigate the impact of this choice on our results, we restrict the $k$ range used in the fit of $\Pcross$ to \kmaxtwo\ , while we fit $\Plya$ up to \kmaxthree. We further check that fitting \Plya\ only and fitting \Pcross\ only yield consistent best-fit values for the parameters shared between the two models, namely $b_\mathrm{F}, b_\eta$ \czeroF, \ctwoF\ and \cfourF. \autoref{fig:bestfits_auto_alone_vs_cross_alone} shows the best-fit values obtained from the two fits as a function of redshift. The linear bias parameters $b_\mathrm{F}$ and $b_\eta$ are consistent within $\sim 1\sigma$ over the full redshift range. By contrast, the counterterms are only weakly constrained by the cross-power spectrum alone, although their best-fit values remain compatible with those inferred from \Plya. This reduced constraining power is partly due to the lower value of $k_\mathrm{max}$ adopted for \Pcross\ relative to \Plya. In addition, the cross-power spectrum model contains a larger number of free parameters, including three additional counterterms associated with the dark matter field. Moreover, in \autoref{fig:triangle_plot_auto_alone_vs_cross_alone_vs_combined} we show the marginalised posterior distributions and two-dimensional contours obtained from fitting \Plya\ only, \Pcross\ only, and from the joint analysis. The best-fit values of all parameters are consistent across the three fits. We further find that the addition of $\Pcross$ to $\Plya$ in the combined analysis does not change the degeneracies between the parameters.

\subsection{Linear theory bias comparison with previous results}

\begin{figure}[tbp]
\centering 
\includegraphics[width=\textwidth]{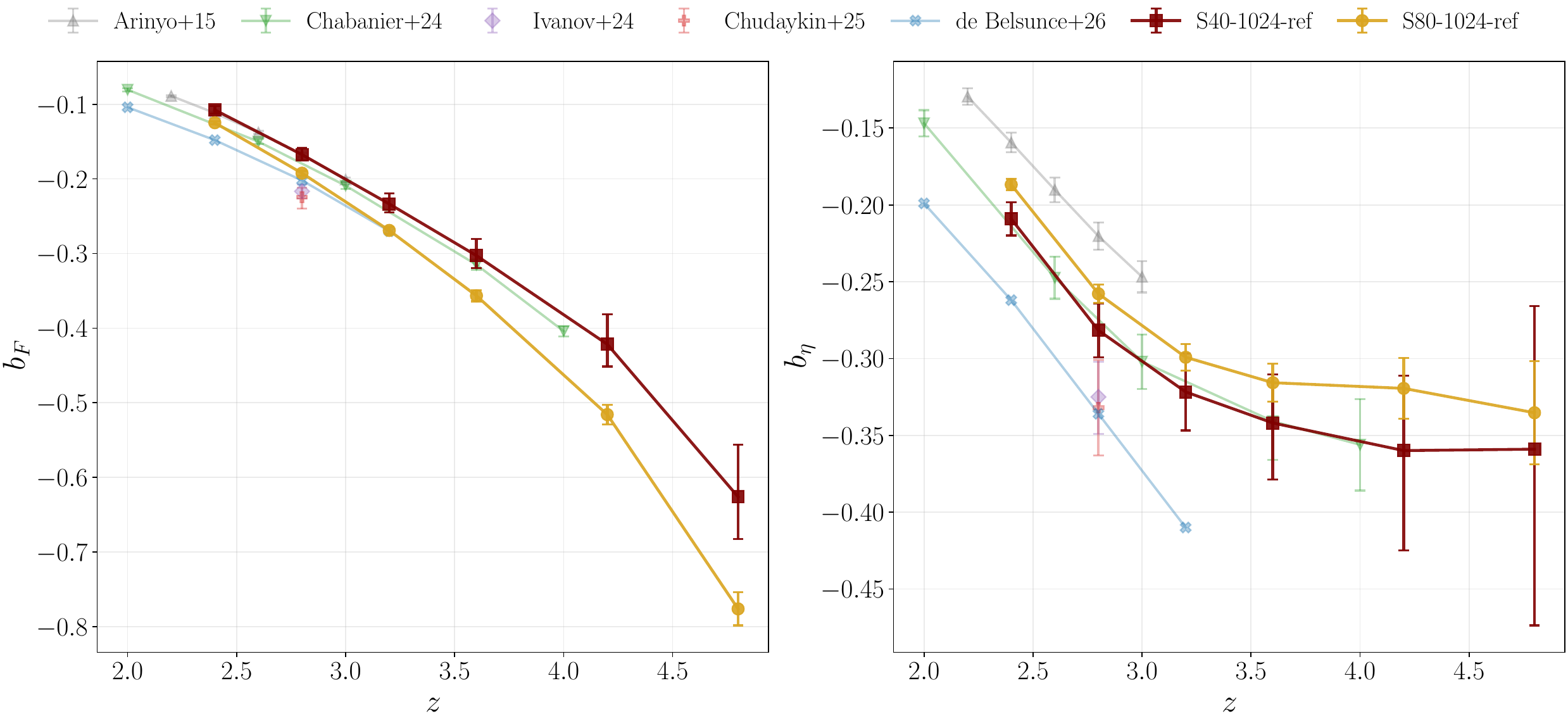}
\hfill
\caption{\label{fig:z_evolution_biases}
Redshift evolution for the linear bias parameters $b_\mathrm{F}$ (left panel) and $b_\eta$ (right panel). We compare this work's results for the S40-1024-ref (maroon) and S80-1024-ref (gold) simulations with results from \cite{Arinyo-i-Prats:2015vqa} (grey), \cite{Chabanier:2024knr} (green), \cite{Ivanov:2023yla} (purple), \cite{Chudaykin:2025gsh} (red), \cite{deBelsunce:2026tks} (blue).}
\end{figure}

In \autoref{fig:z_evolution_biases} we show the redshift evolution of the linear bias parameters, $b_\mathrm{F}$ and $b_\eta$, obtained from joint fits of \Plya\ and \Pcross\ at different redshifts. We show results from this work obtained from the S40-1024-ref and S80-1024-ref simulations in comparison with previous literature results from \cite{Arinyo-i-Prats:2015vqa, Chabanier:2024knr, Ivanov:2023yla, Chudaykin:2025gsh, deBelsunce:2026tks}. For \cite{Arinyo-i-Prats:2015vqa} and \cite{Chabanier:2024knr}, we consider the \texttt{Fiducial} and \texttt{L160R25} simulations respectively.
First, we note that all results exhibit a broadly similar redshift evolution of $b_\mathrm{F}$ and $b_\eta$. The results from the S80-1024-ref simulation show a slightly steeper evolution of $b_\mathrm{F}$ over the range $2.4\leq z\leq 3.2$ compared to the other results, including those obtained from the S40-1024-ref simulation. At higher redshift, $z\gtrsim 3$, the redshift evolution of $b_\mathrm{F}$ exhibited by this work's results is in very good agreement with the results of \cite{Chabanier:2024knr}. On the other hand, we find that for both S40-1024-ref and S80-1024-ref the redshift dependence of $b_\eta$ becomes progressively weaker and tends to flatten. By contrast, the results of \cite{deBelsunce:2026tks} in the range $2.8\leq z\leq 3.2$ do not show the same behaviour, but instead remain consistent with an approximately constant slope. 

For $b_\mathrm{F}$, the best-fit values from the S40-1024-ref simulation agree well with \cite{Arinyo-i-Prats:2015vqa} and \cite{Chabanier:2024knr} over the full redshift range, while at $z=2.8$ they differ from the results of \cite{deBelsunce:2026tks}, \cite{Ivanov:2023yla} and \cite{Chudaykin:2025gsh} by $\sim 20\%, 29\%$ and $34\%$, respectively. The results from the S80-1024-ref simulation are consistent with those from S40-1024-ref and with \cite{Arinyo-i-Prats:2015vqa} and \cite{Chabanier:2024knr} at $z=2.4$. At higher redshift, however, the steeper evolution noted above drives $b_\mathrm{F}$ to more negative values more rapidly, so that the difference with \cite{Arinyo-i-Prats:2015vqa} reaches $\sim 13\%$ at $z=2.8$, and the difference with \cite{Chabanier:2024knr} reaches $\sim 12\%$ at $z=3.6$. We find, however, that S80-1024-ref at $z=2.8$ agrees better than S40-1024-ref with \cite{deBelsunce:2026tks}, \cite{Ivanov:2023yla} and \cite{Chudaykin:2025gsh}, with residual discrepancies of only $\sim 5\%, 13\%$ and $17\%$, respectively.

For $b_\eta$, we find that the results from S40-1024-ref are in very good agreement with \cite{Chabanier:2024knr}, but differ from \cite{Arinyo-i-Prats:2015vqa} by $\sim 22\%$ at $z=2.4$ and $z=2.8$. The S80-1024-ref results differ from both S40-1024-ref and \cite{Chabanier:2024knr} results by $\sim 12\%$ at $z=2.4$ and $\sim 9\%$ at $z=2.8$, and from \cite{Arinyo-i-Prats:2015vqa} by $\sim 15\%$ at the same two redshifts. Both results from our simulations disagree significantly with the results of \cite{Ivanov:2023yla}, \cite{Chudaykin:2025gsh} and \cite{deBelsunce:2026tks}: at $z=2.8$, the discrepancies amount to $\sim 15\%, 18\%$ and $19\%$ for S40-1024-ref and $\sim 26\%, 29\%$ and $30\%$ for S80-1024-ref. 

\subsection{$P_\mathrm{shot}=0$ analysis}

\begin{figure}[tbp]
\centering 
\includegraphics[width=0.9\textwidth]{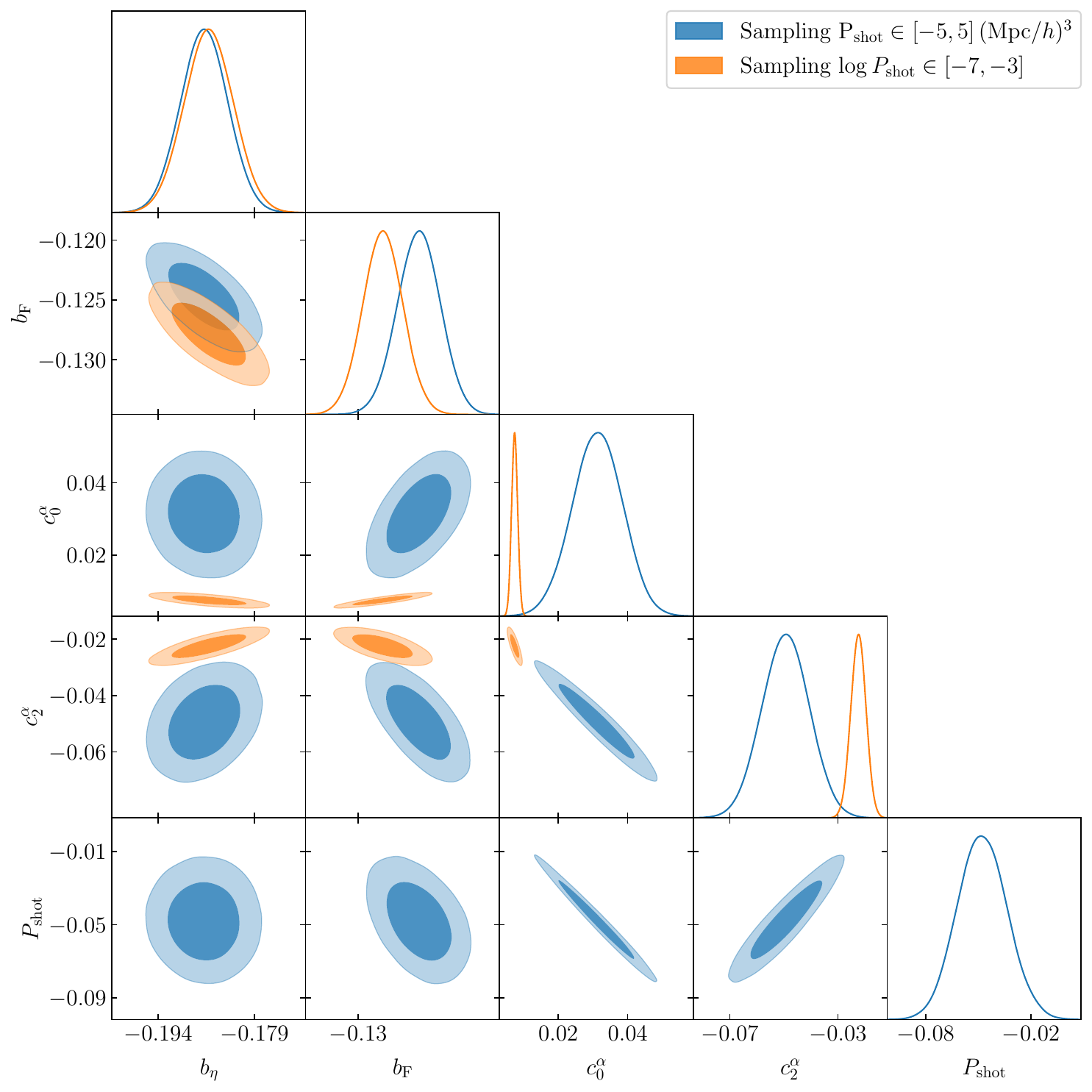}
\hfill
\caption{\label{fig:logPshot_vs_Pshot_triangle_plot}
Triangle plot comparing sampling $P_\mathrm{shot}$ with a uniform prior or sampling $\log_{10}{P_\mathrm{shot}}$ with a uniform prior. We do not show the $P_\mathrm{shot}$ posterior for the latter case as it takes very small values. The results are from fits to the S80-1024-ref simulation. }
\end{figure}

\begin{figure}[tbp]
\centering 
\includegraphics[width=\textwidth]{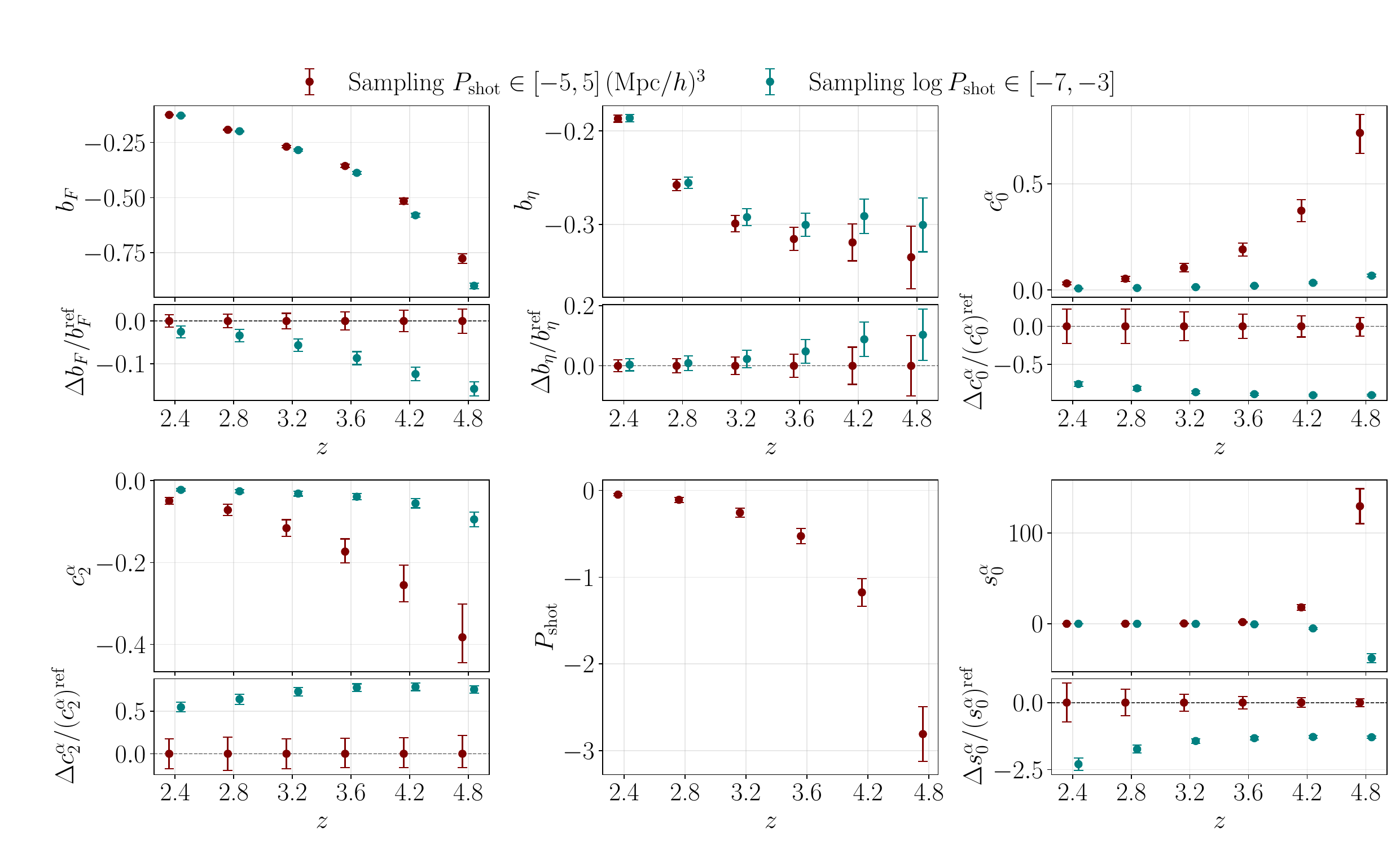}
\hfill
\caption{\label{fig:logPshot_vs_Pshot_bestfits}
Comparison of best-fit values for $b_\mathrm{F}, b_\eta, \czeroF, \ctwoF, P_\mathrm{shot}, s_0^\alpha$ when sampling $P_\mathrm{shot}$ with a uniform prior or sampling $\log_{10}{P_\mathrm{shot}}$ with a uniform prior. The lower panels show the difference of each fit with respect to a \textit{reference} value of each parameter, taken to be the best-fit result of the fit with $P_\mathrm{shot}$ sampled with a uniform prior. The results are obtained from the S80-1024-ref simulation.
}
\end{figure}

\begin{figure}[tbp]
\centering 
\includegraphics[width=\textwidth]{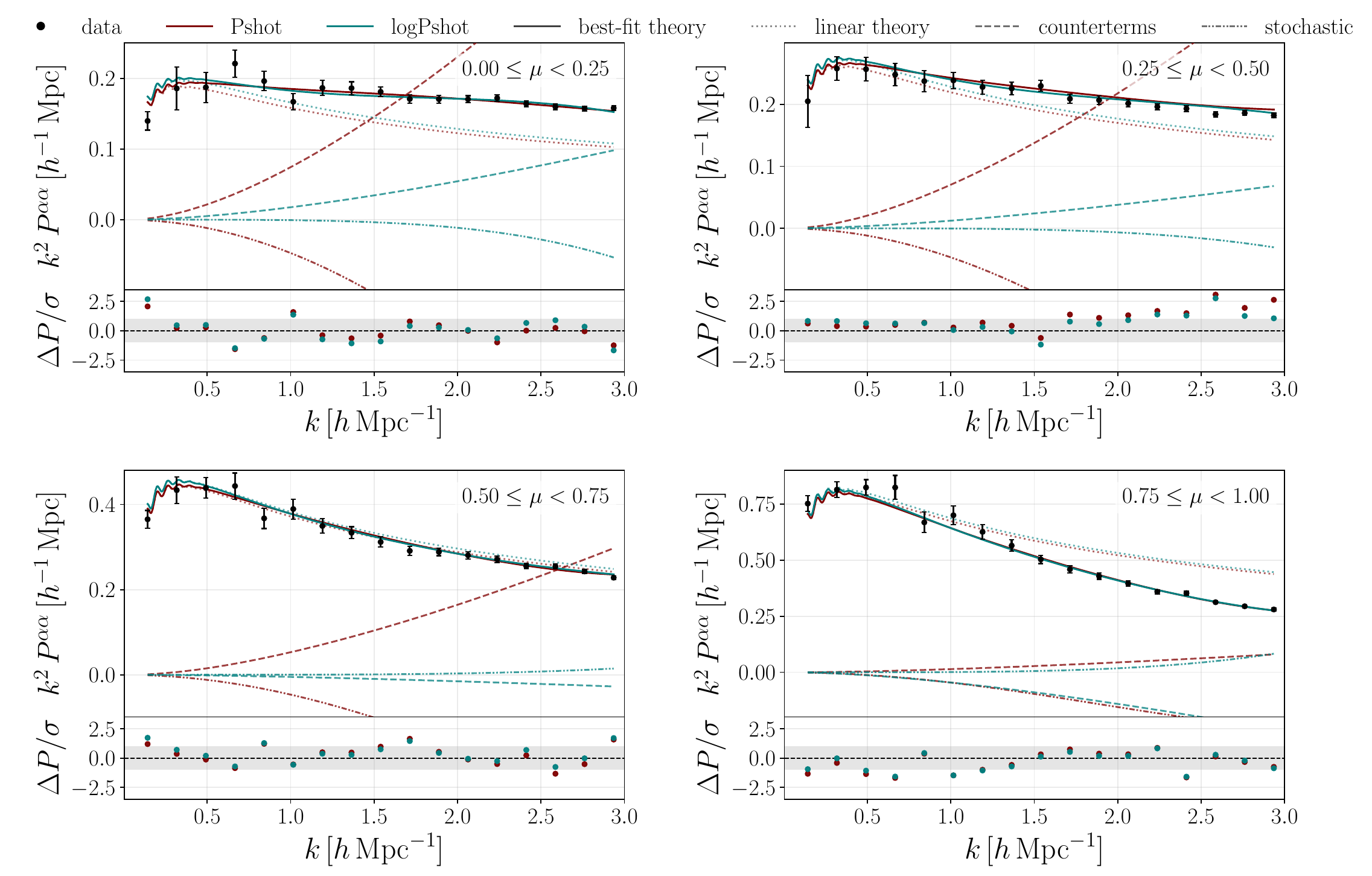}
\hfill
\caption{\label{fig:Pshot_vs_logPshot_bestfit_decomposed}
Comparison of best-fit theory \Plya\ when sampling $P_\mathrm{shot}\in [-5, 5]\; (\mathrm{Mpc}/h)^3$ to the best-fit theory for $\log_{10}{P_\mathrm{shot}}\in[-7,-3]$.
Moreover we plot the linear theory part of the power spectrum in dotted, the counterterms contribution in dashed and the stochastic contribution in dotted-dashed. These results are obtained for the S80-1024-ref simulation.}
\end{figure}

We now turn to examine in detail the role of the shot-noise parameter, $P_\mathrm{shot}$. The treatment of this parameter varies in previous literature works. In \cite{Ivanov:2023yla} and \cite{Chudaykin:2025gsh}, the authors find that the contribution of the shot-noise parameter is negligible and therefore set it to zero in their analysis. In \cite{deBelsunce:2024rvv} instead the authors find a slight preference for a non-zero $P_\mathrm{shot}$ term at low redshifts ($2\lesssim z \lesssim 3$), while they find high-redshift ($3 \lesssim z \lesssim 4$) data to be consistent with a zero shot-noise contribution. Similarly, in \cite{deBelsunce:2025bqc, deBelsunce:2025gci, deBelsunce:2026tks} the authors find that stochastic terms are small. In our minimal EFT framework, we find, in agreement with previous literature, that at low redshifts ($2.4 \leq z \lesssim 3$) the shot-noise term is small. However, we find that the best-fit value of $P_\mathrm{shot}$ tends to shift towards negative, and therefore likely unphysical, values with increasing redshift.

We note that a negative shot-noise contribution is not necessarily unphysical: in \cite{Baldauf:2013hka} the authors noted that halo exclusion on large scales can have a large effect and reduce the effective scale-independent stochastic term $P_\mathrm{s}=P_\mathrm{shot}+P_\mathrm{ex}$, where $P_\mathrm{shot}$ is the usual Poisson contribution and $P_\mathrm{ex}$ is the contribution due to halo exclusion that is expected to be negative. Moreover, they find that in this framework, the leading scale-dependent stochastic contribution, which in the EFT formalism would correspond to $s_0^\alpha$, has a positive contribution to the full stochastic power spectrum. Similarly, in \cite{Irsic:2018hhg} the authors find a similar situation in \lya\ absorbers, noting that absorber exclusion reduces the shot-noise in the 1D power spectrum at large scales. Furthermore, within this framework the shot-noise being more negative at higher redshifts could be explained by noting that at higher redshifts the mean flux is lower, the number of absorbers is higher and their spatial extent, given by the Jeans scale, is also larger, leading to a bigger contribution coming from exclusion. 

While this exclusion picture offers a plausible physical explanation for the trend, we note that the inferred value of $P_\mathrm{shot}$ in our minimal model is also subject to strong degeneracies with other model parameters. We show some of the degeneracies in \autoref{fig:logPshot_vs_Pshot_triangle_plot}, where focusing on $P_\mathrm{shot}$ we see strong degeneracies with the \lya\ counterterms. More specifically, $P_\mathrm{shot}$ is negatively correlated with \czeroF\ and \cfourF, while it is positively correlated with \ctwoF. Moreover, we find that the entire stochastic contribution shows little scale and $\mu$ dependence, suggesting that if exclusion is at play, it is characterised by a very small exclusion radius, $R_{\rm excl}$. Since we fit \Plya\ at much larger scales, the stochastic contribution is dominated by $P_\mathrm{shot}$.

In order to obtain an estimate of the value of $P_\mathrm{shot}$ across the redshift range of our interest, we randomly shuffle the \lyaf\ flux field and compute the power spectrum of the shuffled field. This procedure erases the spatial correlations between pixels, such that the resulting power should come solely from shot-noise and exclusion. We find an estimate of $P_\mathrm{shot}$ that ranges from $10^{-6}$ to $10^{-4}$ (Mpc/$h)^3$, depending on redshift. The reshuffled field can also be fitted with the full stochastic contribution, including $s_0^\alpha$ and $s_2^\alpha$ terms, but this does not significantly improve the fit ($\Delta \chi^2 / \chi^2_{\mathrm{shot}} \simeq 0.12$, where $\chi^2_{\mathrm{shot}}$ refers to the fit with only the constant shot-noise term). We find no evidence for the $s_2^\alpha$ term in the stochastic field, while $s_0^\alpha$ is measured to be non-zero at $\sim 1.2\, \sigma$ at $z=2.4$ and is consistent with zero at $1\sigma$ at $z=4.8$. Even when $s_0^\alpha$ is measured to be non-zero, its value is small, so that the contribution of the $s_0^\alpha k^2/k_\mathrm{NL}^2$ term to the full stochastic power spectrum ranges from $\sim 0.2\%$ at $k\sim 0.5 \,h\,{\rm Mpc}^{-1}$ to $\sim 7\%$ at $k=k_\mathrm{max}=3\,h\,{\rm Mpc}^{-1}$. The redshift evolution of the shot-noise $P_{\rm shot}$ estimated from the reshuffled field is well described by the functional form $P_\mathrm{shot}(z)=P_\mathrm{shot,0}(1+z)^\gamma$. We get the best-fit values $P_\mathrm{shot,0} = (2.357\pm0.011)\times 10^{-8} \,(\mathrm{Mpc}/h)^3$ and $\gamma = 4.563\pm0.003$.

We make use of these results by fitting the data sampling $\log_{10}{P_\mathrm{shot}}$ with a uniform prior $\log_{10}{P_\mathrm{shot}}\in [-7, -3]$, and compare the results to the fits performed with a uniform prior on $P_\mathrm{shot}$. \autoref{fig:logPshot_vs_Pshot_bestfits} shows a comparison of the best-fit values obtained from these two analyses on the S80-1024-ref simulation. Using this informative prior on $\log_{10}{P_\mathrm{shot}}$ removes the strong degeneracies between $P_{\rm shot}$ and the counterterms, resulting in more stable counterterms values across the full redshift range. The best-fit value of \czeroF\ decreases by a factor $\sim 5$ at $z=2.4$ and $2.8$, with this factor increasing to $\sim 10$ at $z\geq 3.2$. A similar trend is observed for \ctwoF, whose best-fit values decrease by a factor of $\sim 3$ at $z=2.4$ and $2.8$, and by up to a factor of $\sim 5$ at the highest redshifts. For \cfourF, the best-fit values are reduced by a factor of $\sim 10$ at $z\leq 3.2$, while the reduction becomes less pronounced at higher redshift, reaching a factor of $\sim 5$ at $z=4.8$. We further note that when sampling $\log_{10}{P_\mathrm{shot}}\in [-7,-3]$ we find no constraining power on $\log_{10}{P_\mathrm{shot}}$ and the posterior is dominated by the prior volume. We have additionally tried to widen the prior on $\log_{10}{P_\mathrm{shot}}$ to $\log_{10}{P_\mathrm{shot}}\in [-10,0]$. We find that the posterior is still prior-dominated, with no impact on the ability to constrain the bias parameters $b_\mathrm{F},b_\eta$. 

As stated above, in \autoref{fig:logPshot_vs_Pshot_triangle_plot} we show 1D marginalised posteriors and 2D contours for the bias parameters, the counterterms and the shot-noise parameter, comparing the two different priors on the shot-noise.
We further note that, in most cases, sampling $\log_{10}{P_\mathrm{shot}}$ primarily causes the best-fit values to move along the degeneracy directions. Considering the counterterms \czeroF, \ctwoF\ and \cfourF, their best-fit values shift toward zero, as noted above, following their respective degeneracies with $P_\mathrm{shot}$. At the same time, the degeneracy directions among the counterterms remain largely unchanged: \czeroF\ and \ctwoF\ remain strongly anticorrelated, as do \ctwoF\ and \cfourF, while \czeroF\ and \cfourF\ remain strongly positively correlated. 
Turning to the bias parameters $b_\mathrm{F}, b_\eta$, both their inferred best-fit values and the $b_\mathrm{F}- b_\eta$ contour remain essentially unchanged. We observe only a small shift of the best-fit values along the degeneracy directions involving the linear biases, $P_\mathrm{shot}$ and counterterms. Although the degeneracy of $b_\mathrm{F}$ and $b_\eta$ with $P_\mathrm{shot}$ is relatively mild, the bias parameters are also correlated with the counterterms. Consequently, when sampling $\log_{10}{P_\mathrm{shot}}$, the counterterms shift along their degeneracy directions with $P_\mathrm{shot}$, inducing the corresponding small shift in $b_\mathrm{F}$ and $b_\eta$. 

Moreover, in \autoref{fig:Pshot_vs_logPshot_bestfit_decomposed} we show the best-fit theory for \Plya, comparing the two priors on $P_\mathrm{shot}$ and we also show the separate contributions of the linear part, counterterms and stochastic terms to the full power spectrum at $z=2.4$. At this redshift, the linear contributions in the two cases are very similar. The contributions from the counterterms and stochastic terms, however, are substantially larger when no prior is placed on $P_\mathrm{shot}$: individually, they tend to be dominant over the linear theory term, but their sum is not, due to strong cancellations between them. On the contrary, when sampling $\log _{10}{P_\mathrm{shot}}$ these cancellations do not occur, and both the counterterms and stochastic contributions remain subdominant compared to the linear theory across the $\mu$ and $k$ range of our fits, suggesting that, in this case, the deviation from linear theory is small. 

\subsection{Effect of mean field normalisation} 

\begin{figure}[tbp]
\centering 
\includegraphics[width=\textwidth]{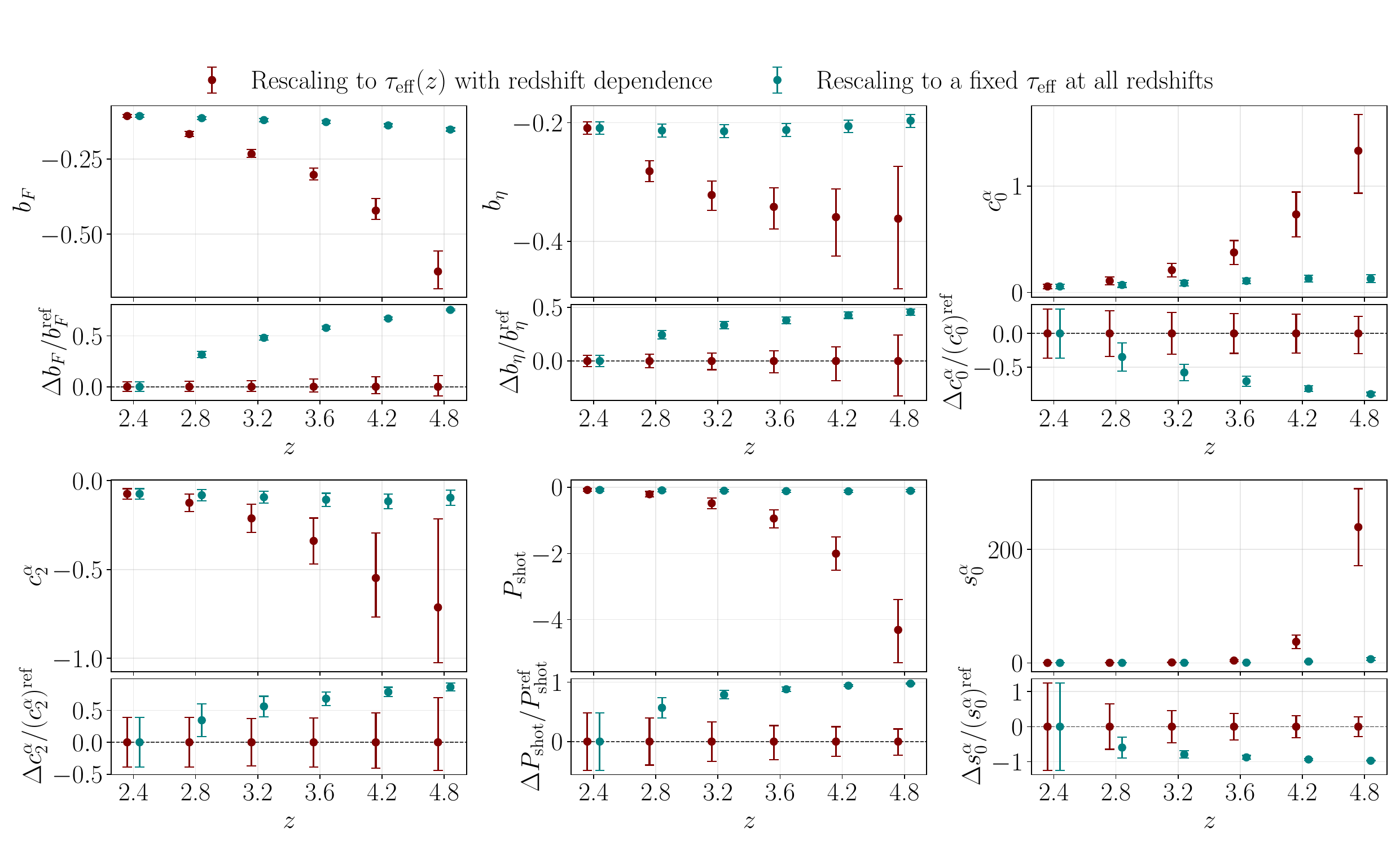}
\hfill
\caption{\label{fig:bestfits_no_tau_evo}
Comparison of best-fit values for $b_\mathrm{F}, b_\eta, \czeroF, \ctwoF, P_\mathrm{shot}, s_0^\alpha$ for the reference case to when rescaling the mean flux to the same value at all redshifts, taking away the redshift evolution of $\tau_\mathrm{eff}$. We compare fits when sampling $P_\mathrm{shot}\in [-5, 5]\;(\mathrm{Mpc}/h)^3$ for the S40-1024-ref simulation.}
\end{figure}

In the previous section we have observed that when sampling $\log_{10}{P_\mathrm{shot}}$, thus keeping $P_\mathrm{shot}$ small, we find small deviations from linear theory. 
To investigate this further, we recompute the power spectra for the S40-1024-ref simulation by rescaling the mean flux to the same value, taken from $z=2.4$, at all redshifts. This removes the redshift evolution of $\tau_\mathrm{eff}$, which is expected to be the dominant driver of the redshift evolution of the linear biases $b_\mathrm{F}, b_\eta$ and counterterms \czeroF, \ctwoF, \cfourF\ over the range of scales of our interest.

As previously discussed, in the IGM that produces the \lyaf, the hydrogen is highly photoionised and approximately in equilibrium with the ionising UV background, therefore the neutral hydrogen density is determined by the balance between photoionisation, occurring at rate $\Gamma_{\mathrm{HI}}$, and recombination, whose coefficient $\alpha(T)$ at the temperatures of relevance for the forest scales as $\alpha\propto T^{-0.7}$. Consequently, the optical depth $\tau$ scales as $\tau \propto \Delta_\mathrm{b}^2 T^{-0.7}/\Gamma_\mathrm{HI}$, where $\Delta_\mathrm{b}=1+\delta_\mathrm{b}$ and $\delta_\mathrm{b}$ denotes the gas overdensity. Since $\tau\propto\Gamma_{\rm HI}^{-1}$, rescaling $\tau_\mathrm{eff}$, for fixed temperature and density field, is equivalent to rescaling the photoionisation rate. This rescaling, however, does not account for the full redshift evolution of the \lya\ biases $b_\mathrm{F}, b_\eta$. Their evolution also receives contributions from changes in the gas temperature $T$, which modifies the mapping between the gas density and the transmitted flux. Additionally, on sufficiently large scales the gas overdensity traces the underlying matter overdensity, $\delta_\mathrm{b} \propto \delta_m$, while in linear theory $\delta_m = D(z)\delta_{m,0}$, with $D(z)$ the linear growth factor. Because the transmitted \lyaf\ flux is a non-linear function of the underlying density field, the bias parameters $b_\mathrm{F}$ and $b_\eta$ can also get a dependence on the growth of matter perturbations. In \autoref{fig:bestfits_no_tau_evo} we compare the best-fit values for the biases and counterterms obtained from the reference simulations to those in which the $\tau_\mathrm{eff}$ evolution has been removed. As expected, $b_\mathrm{F}$ and $b_\eta$ retain a residual redshift evolution, and some redshift evolution is also present in the counterterms and stochastic terms. Interestingly, removing the evolution of $\tau_\mathrm{eff}$ shifts the best-fit counterterms towards zero. This suggests that the evolution of $\tau_\mathrm{eff}$ is the main driver of their redshift dependence but, over the scales considered here, there is evidence for redshift evolution of the model parameters beyond the evolution of $\tau_\mathrm{eff}$, suggesting that a single parameter is not sufficient to describe the linear bias parameters $b_\mathrm{F}, b_\eta$ over the full redshift range.

\subsection{The effects of box size and resolution}
\begin{figure}[tbp]
\centering 
\includegraphics[width=\textwidth]{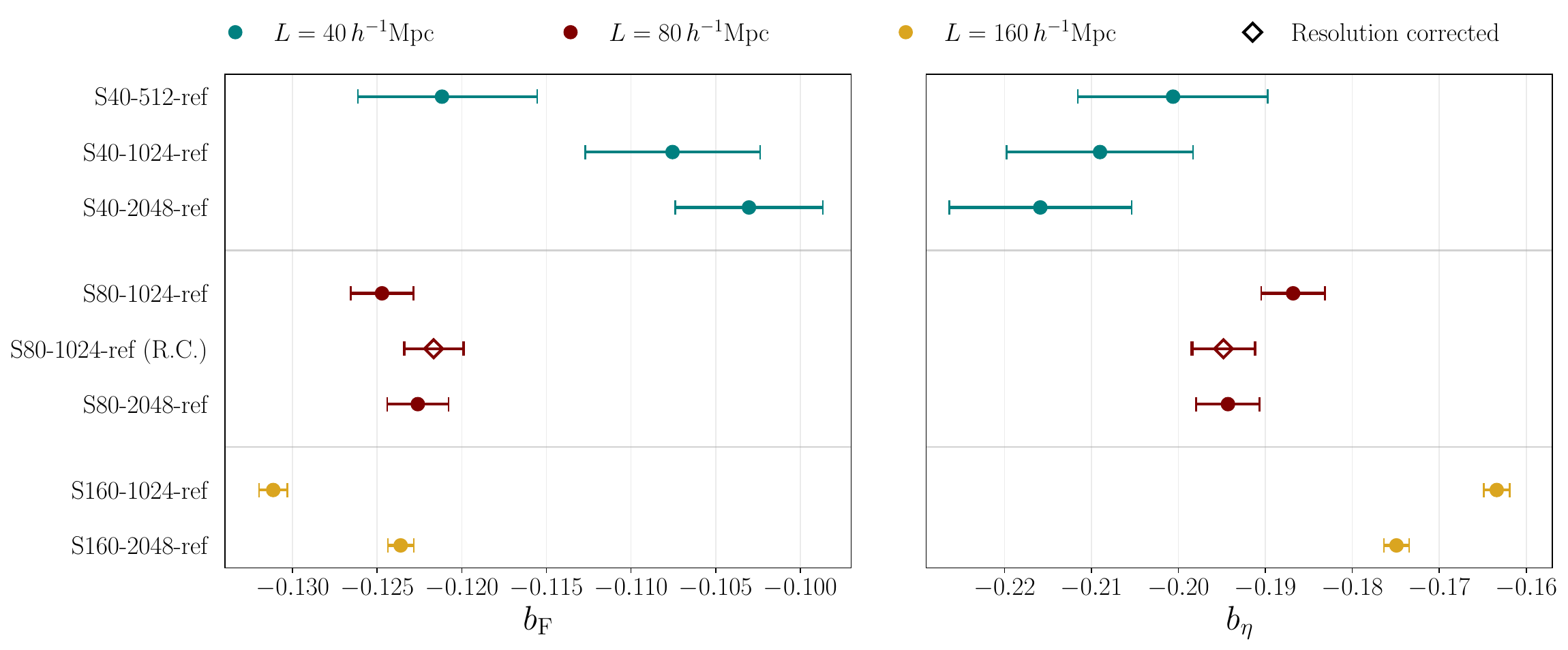}
\hfill
\caption{\label{fig:diff_sim_biases}
Comparison of best-fit results for $b_\mathrm{F}$ and $b_\eta$ for multiple simulations with different box size and resolution, at $z=2.4$.}

\end{figure}

In \autoref{fig:diff_sim_biases} we show the best-fit values of the linear bias parameters $b_\mathrm{F}$ and $b_\eta$ obtained from simulations spanning different box sizes and resolutions. We find that changing the box size at fixed resolution has a small effect on $b_\mathrm{F}$, whereas $b_\eta$ changes by approximately $7\%$ across the three box sizes considered, $L_\mathrm{box}=40, 80, 160\,h^{-1}\,{\rm Mpc}$. Moreover, the box-size dependence of $b_\eta$ does not appear to converge over the range of box sizes explored, suggesting that larger simulation volumes may be required and that current estimates of $b_\eta$ include some box size dependence.

We next assess the impact of the simulation's resolution by comparing simulations with the same box size but different number of particles. For simulations with $L_\mathrm{box}=40\,h^{-1}\,{\rm Mpc}$ (in teal in the plot), increasing the number of gas and dark matter particles from $512^3$ to $1024^3$ changes $b_\mathrm{F}$ by approximately $12\%$ and $b_\eta$ by approximately $4\%$. A further increase from $1024^3$ to $2048^3$ produces shifts of about $4\%$ in $b_\mathrm{F}$ and $3\%$ in $b_\eta$. For $L_\mathrm{box}=80\,h^{-1}\,{\rm Mpc}$ (in maroon), the difference when increasing the number of particles from $1024^3$ to $2048^3$ is approximately $2\%$ for $b_\mathrm{F}$ and $4\%$ for $b_\eta$, while for $L_\mathrm{box}=160\,h^{-1}\,{\rm Mpc}$ (in gold) we find shifts of approximately $5\%$ and $7\%$, respectively. The $L_\mathrm{box}=40\,h^{-1}\,{\rm Mpc}$ simulations, which span the widest range in resolution, show evidence of convergence for $b_\mathrm{F}$, while the convergence of $b_\eta$ is more modest. Moreover, we find that all counterterms and stochastic terms also show resolution and box size effects that are included in the best-fit values.

We note that correcting for box size is generally more difficult than correcting for resolution, since box-size corrections are estimated from the large-volume simulation and are therefore strongly affected by its sample variance. A more robust strategy is instead to combine large-volume, lower-resolution simulations with smaller-volume, higher-resolution runs, using the latter to calibrate the resolution correction at substantially lower computational cost. To test this approach, we applied the splicing technique of \cite{McDonald:2001fe} to the S80-1024-ref simulation, using the S40-512-ref and S40-1024-ref simulations to estimate the resolution correction needed to match the effective resolution of the S80-2048-ref run. The splicing procedure performs remarkably well: it shifts the inferred values of both $b_\mathrm{F}$ and $b_\eta$ towards those obtained directly from the S80-2048-ref simulation, reducing the discrepancy in $b_\mathrm{F}$ from $\sim 2\%$ to $\sim 0.3\%$ and in $b_\eta$ from $\sim4\%$ to $\sim0.7\%$. Similarly, the residual percentage differences in counterterms and stochastic terms decrease by factors of $\sim 4$ and $\sim 5$, respectively. These results confirm that splicing provides an accurate and computationally efficient method to correct for resolution effects. 

\subsection{Reionisation history dependence of bias parameters}

\begin{figure}[tbp]
\centering 
\includegraphics[width=\textwidth]{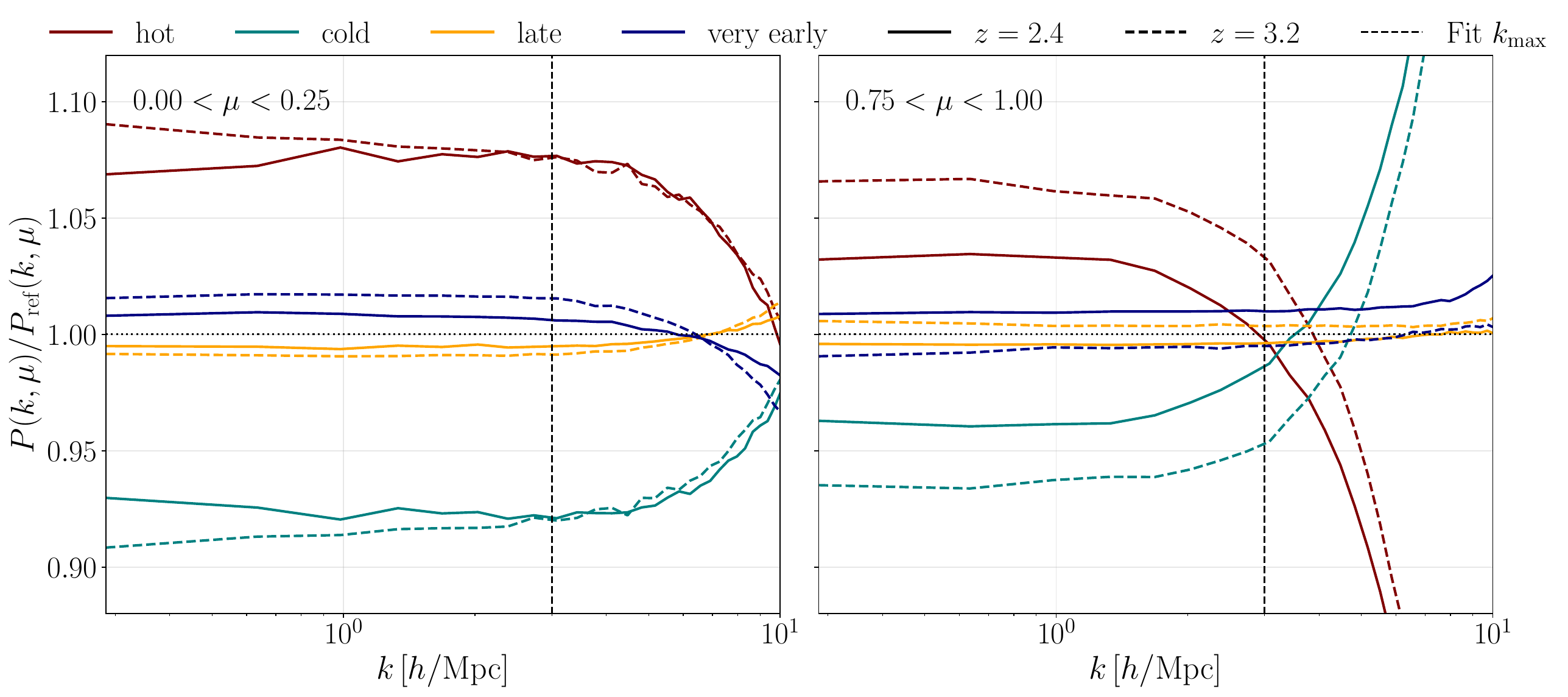}
\hfill
\caption{\label{fig:relics_spectra_comp}
Comparison of the \lyaf\ auto-power spectra of Sherwood--Relics simulations with $L_\mathrm{box}=40\,h^{-1}\,{\rm Mpc}$ for different reionisation histories. We show the ratio to the R40-1024-ref simulation up to $k=10\,h\,{\rm Mpc}^{-1}$. Solid lines represent the ratios for $z=2.4$, dashed lines for $z=3.2$ instead. The dashed vertical line indicates the $k_\mathrm{max}$ of the fit. We omit the \textit{early} reionisation model as its ratio with the \textit{ref} is close to $1$. 
}
\end{figure}

We now examine how the parameters depend on reionisation and thermal history, analysing fits to Sherwood--Relics simulations run with different thermal and reionisation histories. In \autoref{fig:relics_spectra_comp} we show the ratios of \Plya\ measured from simulations with $L_\mathrm{box} =40\,h^{-1}\,{\rm Mpc}$ and $N_\mathrm{part}=2\times1024^3$, with varying reionisation histories to those of the reference simulation with $z_\mathrm{rei}^{\mathrm{end}}=6.0$. We focus on scales up to \kmaxthree, the smallest scale included in our fits. Changing $z_\mathrm{rei}^{\mathrm{end}}$ to $z_\mathrm{rei}^{\mathrm{end}}\simeq5.37, 6.7, 7.4$ produces changes smaller than $2\%$ in both \Plya\ and \Pcross\ across all $\mu$-bins, at both $z=2.4$ and $z=3.2$. We therefore expect no significant variation in the best-fit parameters across these data sets. 
By contrast, the \textit{hot} and \textit{cold} reionisation models show larger deviations from the reference simulation, reaching $\sim 10\%$ in \Plya\ and $\sim 5\%$ in \Pcross. The ratios for these models also exhibit a $\mu$ dependence, with deviations decreasing toward higher $\mu$ at low redshift ($2.4\leq z\lesssim 3.6$) and increasing with $\mu$ at higher redshift ($z\gtrsim 3.6$). Over the scales that we include in the analysis, however, the differences remain approximately scale independent, therefore we expect the best-fit values for the linear biases $b_\mathrm{F}, b_\eta$ to shift significantly for the \textit{hot} and \textit{cold} models, while the scale- and $\mu$- dependent counterterms and stochastic terms remain largely unchanged. 

\begin{figure}[tbp]
\centering 
\includegraphics[width=\textwidth]{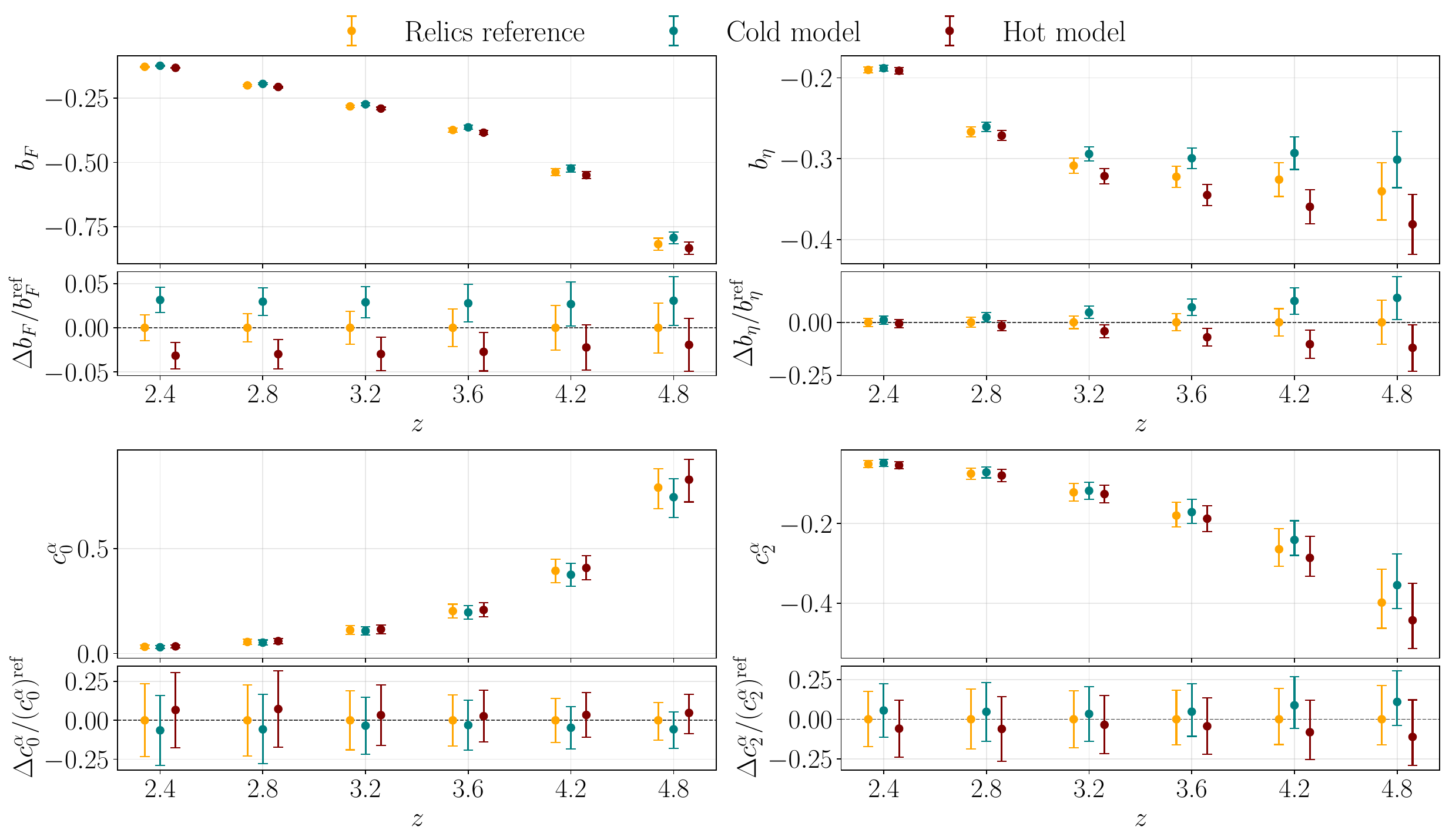}
\hfill
\caption{\label{fig:diff_reio_bestfits_80}
Best-fit parameters from combined fits of \Plya\ and \Pcross\ at $z=2.4, 2.8,3.2,3.6,4.2,4.8$ for the R80-1024-ref, R80-1024-hot and R80-1024-cold simulations. In the lower panels we show the difference between the best fits for the different reionisation histories, normalised by the \textit{reference} value.
}
\end{figure}

In \autoref{fig:diff_reio_bestfits_80} we show the best-fit values for $b_{\mathrm{F}}, b_\eta, \czeroF$ and $\ctwoF$ obtained from jointly fitting \Plya\ and \Pcross\ measured from the R80-1024-ref, R80-1024-hot and R80-1024-cold simulations, at varying redshift. For $b_{\mathrm{F}}$, the \textit{hot} and \textit{cold} models exhibit comparable deviations from the \textit{reference} model, remaining approximately constant at $\sim 3\%$ across the entire redshift range. On the contrary, the difference in $b_\eta$ best-fit values increases with redshift, from $\sim 1\%$ at $z=2.4$ to $\sim 11\%$ at $z=4.8$. This can be understood from the $\mu$-dependence of the ratios of the \textit{hot} and \textit{cold} model spectra relative to the \textit{reference} ones, shown in \autoref{fig:relics_spectra_comp}. At low redshifts ($z=2.4,\, 2.8$), the ratios are approximately flat or slightly decreasing with $\mu$, providing little sensitivity to $b_\eta$. At higher redshifts, however, this trend reverses, with ratios increasing with $\mu$, thereby increasing the sensitivity to $b_\eta$ and allowing us to distinguish between different values across reionisation models.
For counterterms and stochastic terms we find that the best-fit values of the \textit{hot} and \textit{cold} models are consistent with the \textit{reference} model within $1\sigma$ across the full redshift range. Moreover, we note that, other than increasing the simulation volume, the primary way to improve the sensitivity to both the linear bias parameters and the scale-dependent model parameter is to extend the fitted scale range toward smaller scales. This is, however, limited by the regime of validity of the EFT framework, and extending the theoretical model to one-loop order does not help in overcoming this limitation.

\subsection{Empirical correlations with the \lya\ bias}
\begin{figure}[tbp]
\centering 
\includegraphics[width=\textwidth]{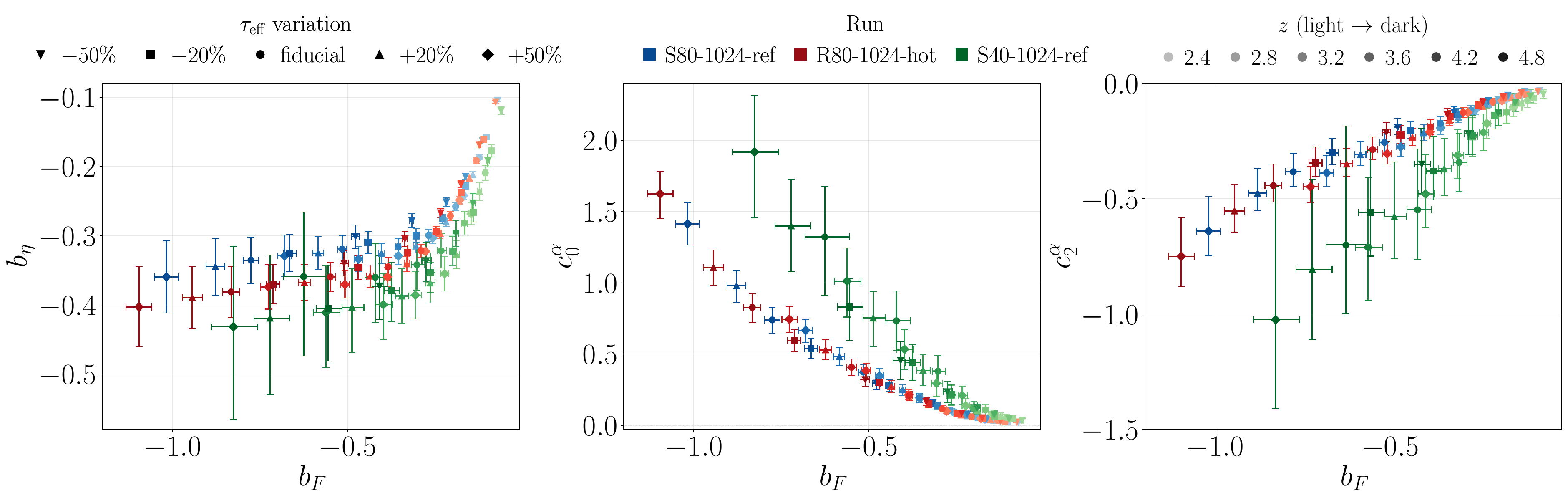}
\hfill
\caption{\label{fig:b_eta_and_c0_and_c2_vs_bf}
Evolution of $b_\eta$ (left panel), $\czeroF$ (central panel) and $\ctwoF$ (right panel) across different redshifts and $\tau_\mathrm{eff}$ values for the S40-1024-ref, S80-1024-ref and R80-1024-hot simulations. Each simulation is assigned a different colour, with shading indicating redshift, while different markers denote the different $\tau_\mathrm{eff}$ values.
}
\end{figure}

\begin{figure}[tbp]
\centering 
\includegraphics[width=\textwidth]{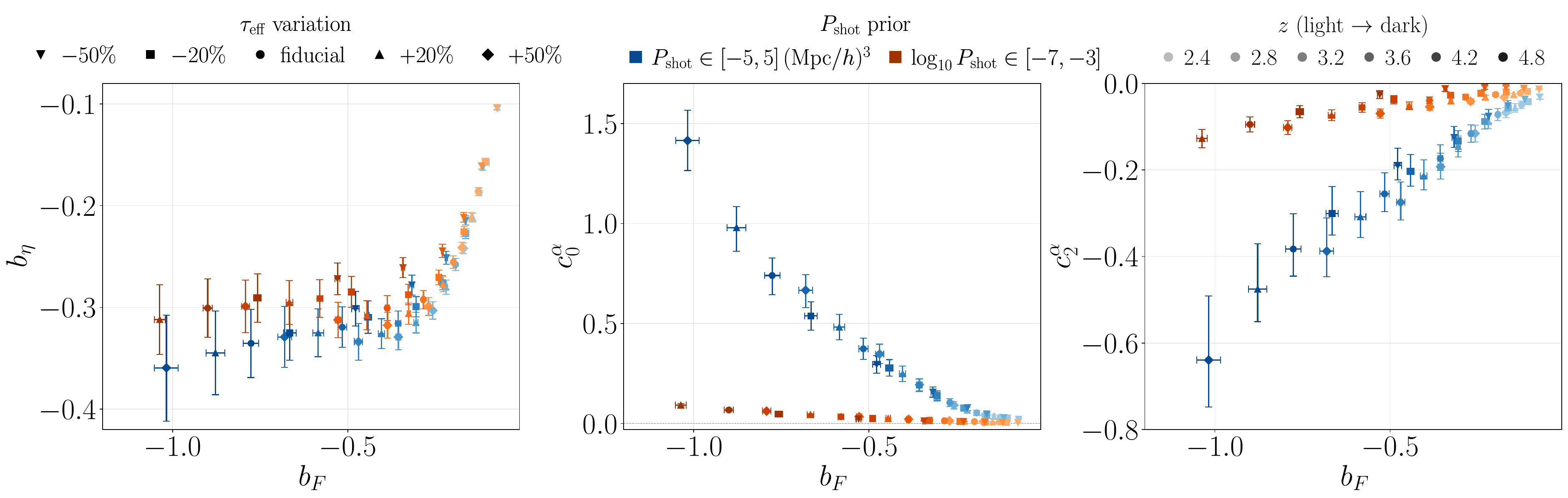}
\hfill
\caption{\label{fig:b_eta_and_c0_and_c2_vs_bf_Pshot_vs_logPshot}
Evolution of $b_\eta$ (left panel), $\czeroF$ (central panel) and $\ctwoF$ (right panel) across different redshifts and $\tau_\mathrm{eff}$ values for the S80-1024-ref simulation, comparing fitting with $P_\mathrm{shot}\in [-5, 5]\,(\mathrm{Mpc}/h)^3$ to fitting with $\log_{10}{P_\mathrm{shot}}\in[-7,-3]$.
}
\end{figure}
We turn our attention to the dependence of EFT model parameters on the linear density bias $b_\mathrm{F}$ through empirical correlations. We look at three simulations: S40-1024-ref, S80-1024-ref and R80-1024-hot, and for each simulation we look at redshifts $z=2.4,2.8,3.2,3.6,4.2,4.8$. We also consider the evolution of parameters with $\tau_\mathrm{eff}$ in the following way: as described above, at each redshift the mean optical depth is rescaled to match the observed value and redshift evolution as found in \cite{Viel:2013fqw}. Then for each redshift, we also rescale the data to a value of $\tau_\mathrm{eff}$ that is shifted by $20\%$ and $50\%$ up and down compared to the observed value, effectively creating multiple data sets that allow us to study the evolution of the parameters with $\tau_\mathrm{eff}$ across redshifts. In \autoref{fig:b_eta_and_c0_and_c2_vs_bf} we show the redshift and $\tau_\mathrm{eff}$ evolution of $b_\eta$, $\czeroF$ and $\ctwoF$ plotted against $b_\mathrm{F}$, for the three simulations. We find a significant effect of box size and resolution, which primarily affects the relations between the counterterms $\czeroF$ and $\ctwoF$, and $b_\mathrm{F}$, while it is less apparent in the correlation between $b_\eta$ and $b_\mathrm{F}$. Furthermore, the relations show non-negligible scatter, since the data, over the range of scales considered, are sensitive not only to $\tau_\mathrm{eff}$ but also to the astrophysical parameters governing the relation between the gas temperature and the baryon overdensity, $T_0$ and $\gamma$, and on the cosmological parameters. Because of this, one single parameter is insufficient to fully describe the counterterms, and this limitation gives rise to the observed scatter.

Moreover, in \autoref{fig:b_eta_and_c0_and_c2_vs_bf_Pshot_vs_logPshot} we compare the correlation between $b_\eta$, $\czeroF$, $\ctwoF$ and $b_\mathrm{F}$ obtained from fits in which we sample $P_\mathrm{shot}\in [-5, 5]\;(\rm{Mpc}/h)^3$ to those obtained when sampling $\log_{10}{P_\mathrm{shot}} \in[-7,-3]$. For this comparison we use the S80-1024-ref simulation. We find no significant difference in the $b_\eta$-$b_\mathrm{F}$ plane between the two analyses, despite the best-fit values of the two parameters differing significantly between them, as discussed in previous sections. In contrast, we find substantial differences in the correlation between the counterterms and $b_\mathrm{F}$: in the case where we sample $\log_{10}{P_\mathrm{shot}}$, the counterterms best-fit values are much closer to zero, particularly at high redshift. As noted previously, this can be understood primarily as a consequence of the degeneracy between the counterterms and the shot-noise term. When sampling $P_\mathrm{shot}\in [-5, 5]\; (\mathrm{Mpc}/h)^3$, the shot-noise best-fit value is driven towards negative values, and since $P_\mathrm{shot}$ and $\czeroF$ are negatively correlated, this in turn drives the best-fit value of $\czeroF$ towards larger positive values. Conversely, since $P_\mathrm{shot}$ and $\ctwoF$ are positively correlated, the value of $\ctwoF$ is pushed towards more negative values. 

We find that the correlations between $b_\eta$, $\czeroF$, $\ctwoF$ and $b_\mathrm{F}$ are well described by a power law, or in the case of $b_\eta$, a broken power law, with some intrinsic scatter. $b_\eta=A[(-b_\mathrm{F})/b_\mathrm{F}^{\rm br}]^{\alpha}$ for
$-b_\mathrm{F}<b_\mathrm{F}^{\rm br}$ and
$b_\eta=A[(-b_\mathrm{F})/b_\mathrm{F}^{\rm br}]^{\beta}$ for
$-b_\mathrm{F}\geq b_\mathrm{F}^{\rm br}$. The uncertainties are taken from the measurement errors on the parameters, to which we add an intrinsic scatter term, so that the total variance is $\sigma^2_\mathrm{tot} = \sigma^2_{\mathrm{meas}} + \sigma^2_{\mathrm{i}}$. For the counterterms $\czeroF$ and $\ctwoF$, the correlation is instead well described by the simple power law, $\czeroF =A(-b_\mathrm{F})^\gamma$ across all values of $b_\mathrm{F}$, and likewise for $\ctwoF$. Here again we include the intrinsic scatter term defined above. The results of these fits are reported in \autoref{tab:empirical_fit_relations}.

\subsection{Linear theory bias comparison with theoretical predictions}

The values of $b_\mathrm{F}$ and $b_\eta$ measured in previous sections can be compared against analytic predictions from models of the Lyman-$\alpha$ forest. Two such frameworks of particular relevance here are the fluctuating Gunn-Peterson model of \cite{seljak2012bias}, and the absorber model of \cite{Irsic:2018hhg}. The first treats the transmitted flux as a local, non-linear transformation of the underlying density field and expresses the linear biases as moments of the flux probability distribution. In this picture $b_\mathrm{F}$ and $b_\eta$ are set entirely by the one-point statistics of $F$, together with the two parameters $A$ and $\alpha$, which encode the amplitude of the optical depth and the slope of the temperature-density relation. The second instead decomposes the forest into a population of discrete absorbers, each characterised by a column density $N_{\rm HI}$, with each line being treated as a biased tracer of the underlying matter fluctuations. The linear bias of each absorber is computed by abundance matching against the linear density field, regulated by the exclusion parameters $\epsilon_<$ and $\epsilon_>$, which account for gas removed from the forest by shock heating and collapse. 

The comparison is most straightforward for $b_\eta$. In the model of \cite{seljak2012bias}, $b_\eta$ is fixed once the flux distribution is specified, and is independent of $A$ and $\alpha$. There is, therefore, no freedom to adjust its normalisation to match our results. Nonetheless, the redshift evolution predicted by that model is close to what we recover from our fits. The absorber model reproduces our measured values well at $z=2.4$ and $z=2.8$, but predicts a substantially steeper evolution towards higher redshift than we observe, with the discrepancy growing to a factor of $\sim 3.5$ at $z=4.8$. For $b_\mathrm{F}$, instead, the model of \cite{seljak2012bias} reproduces our measured amplitudes at $z\lesssim4$ for fixed $A$ and $\alpha$, although both parameters are expected to evolve with redshift in a way that is not well constrained, and this evolution could change the predicted values for $b_\mathrm{F}$ significantly. Within the absorber model, our measurements can be matched by adopting $\epsilon_>$ values that range between $0.15$ and $0.25$ at all redshifts, which are considerably smaller than the fiducial value advocated by \cite{Irsic:2018hhg} where they find $\epsilon_> \sim 0.5$.

Neither framework therefore accounts for both biases simultaneously across our full redshift range with parameters that are independently motivated. A more complete treatment of these models, and of the redshift evolution of their parameters, will be important for understanding how the small-scale physics of the IGM influences the large-scale clustering of the \lya\ forest.

\section{Conclusions}\label{sec:conclusions}

In this work we have studied the three-dimensional \lyaf\ flux power spectrum measured from the Sherwood and Sherwood--Relics hydrodynamical simulations with a simplified EFT description in which we close the expansion at tree-level and retain counterterms and stochastic terms. We considered both the flux auto-power spectrum, \Plya, and the cross-power spectrum between the transmitted flux and the dark matter density field, \Pcross. The goal was to assess how far a reduced EFT-inspired description can be pushed, which parameters are actually required by the data, and how robust the inferred parameters are to changes in redshift, thermal history, box size and resolution.

We find that a tree-level model supplemented by the leading counterterms and stochastic contributions is able to fit the simulated power spectra well over the range of scales used in our analysis, in which we fit the \lyaf\ auto-power spectrum up to \kmaxthree\ and the \lyaf--DM cross-power spectrum up to \kmaxtwo. This makes the framework useful for future analysis of the three-dimensional \lyaf, where an EFT-motivated model can be used to describe the flux power spectrum beyond the purely linear regime. 

At the same time, our results show that parameter degeneracies are important even in this simplified model. In particular, the shot-noise term, the counterterms and the linear bias parameters are strongly correlated. We find that allowing an unconstrained constant stochastic contribution can drive the best-fit value of $P_\mathrm{shot}$ to negative values, especially at higher redshifts, and this in turn shifts the inferred counterterms and, to a lesser extent, the bias parameters. This highlights a practical point for future, real-data analysis: increasing the number of nuisance parameters is not automatically conservative if the data do not constrain them independently.

Moreover, we investigated the redshift evolution of the linear bias parameters and compared our results with previous simulation-based analyses. The overall trends are consistent but the specific values depend on the theoretical model, the fitted range of scales and the simulation properties. We find that the main driver of the redshift evolution of the fitted parameters is the effective optical depth $\tau_\mathrm{eff}$. Variations in the thermal and reionisation history mainly affect the linear biases, while scale-dependent EFT parameters do not show a significant dependence over the scales considered. 
Furthermore, we find that inferred parameters, especially the linear biases and counterterms, show a clear dependence on the simulation resolution and, to a lesser extent, on the box size. However, we find that correcting the resolution of a simulation with the splicing technique drastically reduces these differences. 

We then explored empirical correlations between the fitted parameters and the linear density bias $b_\mathrm{F}$. These relations may provide a useful way to reduce the number of free parameters in future analysis, but they also show intrinsic scatter and residual dependence on the simulation box size and resolution. A robust application to real data will therefore require calibration on converged simulations covering the relevant range of cosmological and astrophysical parameters. 

Finally, we compared our linear biases with the analytic predictions of \cite{seljak2012bias} and \cite{Irsic:2018hhg}. Both models can match our $b_\mathrm{F}$, while neither reproduces $b_\eta$ beyond $z\simeq3$: the absorber model agrees at $z=2.4$ and $z=2.8$ but evolves too steeply thereafter, while \cite{seljak2012bias} matches the shape of the evolution but not its amplitude. Exploring this in future works will be important for connecting the small-scale physics of the IGM to the large-scale clustering of the forest.

\acknowledgments
The authors thank Stephen Chen for providing the {\tt ZeNBu} code, and thank Zvonimir Vlah, Roger de Belsunce and Martin White for useful discussions.
The authors are partly supported by the INFN INDARK grant. MV is also supported by SISSA IDEAS grant, the INAF Theory Grant "Cosmological Investigation of the Cosmic Web", and by the Fondazione ICSC, Spoke 3 Astrophysics and Cosmos Observations, National Recovery and Resilience Plan Project ID CN\_00000013 ``Italian Research Center on High-Performance Computing, Big Data and Quantum Computing'' funded by MUR Missione 4 Componente 2 Investimento 1.4: Potenziamento strutture di ricerca e creazione di "campioni nazionali di R\&S (M4C2-19 )" - Next Generation EU (NGEU).
TŠ also acknowledges the support by INAF Theory grant `Cosmological Investigation of the Cosmic Web' (C93C23006820005).

The simulations used in this work were performed using the Joliot Curie supercomputer at the Très Grand Centre de Calcul (TGCC) and the Cambridge Service for Data Driven Discovery (CSD3), part of which is operated by the University of Cambridge Research Computing on behalf of the STFC DiRAC HPC Facility (www.dirac.ac.uk). We acknowledge the Partnership for Advanced Computing in Europe (PRACE) for awarding us time on Joliot Curie in the 16th call. The DiRAC component of CSD3 was funded by BEIS capital funding via STFC capital grants ST/P002307/1 and ST/R002452/1 and STFC operations grant ST/R00689X/1. This work also used the DiRAC@Durham facility managed by the Institute for Computational Cosmology on behalf of the STFC DiRAC HPC Facility. The equipment was funded by BEIS capital funding via STFC capital grants ST/P002293/1 and ST/R002371/1, Durham University and STFC operations grant ST/R000832/1. DiRAC is part of the National e-Infrastructure. Postprocessing and MCMC chains were performed on the Ulysses supercomputer at SISSA.

\appendix 
\newpage
\section{Table for empirical relation fits}

\begin{table}[!ht]
\centering
\scriptsize
\setlength{\tabcolsep}{2.2pt}
\renewcommand{\arraystretch}{1.15}

\newcommand{\paramcellfive}[1]{%
\parbox[c][5.6\baselineskip][c]{1.65cm}{\centering #1}%
}

\newcommand{\paramcellthree}[1]{%
\parbox[c][3.6\baselineskip][c]{1.65cm}{\centering #1}%
}

\newcommand{\valerr}[3]{#1^{+#2}_{-#3}}
\newcommand{\scierr}[4]{\left(#1^{+#2}_{-#3}\right)\cdot 10^{#4}}

\caption{
Best-fit empirical correlations between EFT parameters and the density bias $b_\mathrm{F}$.
The velocity-bias relation is fitted with a broken power law,
$b_\eta=A[(-b_\mathrm{F})/b_\mathrm{F}^{\rm br}]^{\alpha}$ for
$-b_\mathrm{F}<b_\mathrm{F}^{\rm br}$ and
$b_\eta=A[(-b_\mathrm{F})/b_\mathrm{F}^{\rm br}]^{\beta}$ for
$-b_\mathrm{F}\geq b_\mathrm{F}^{\rm br}$.
The counterterm relations are fitted as $c_i^{\alpha}=A(-b_\mathrm{F})^\gamma$, with $i=0,2$.
Each cell reports the best-fit parameters and the intrinsic scatter entering
$\sigma_{\rm tot}^2=\sigma_{\rm meas}^2+\sigma_{\rm i}^2$. The two columns for each simulation, named $P_\mathrm{shot}$ and $\log_{10}{P_\mathrm{shot}}$, indicate the shot-noise prior adopted when inferring the EFT parameters that enter these fits: $P_{\rm shot}\in [-5,5]\,(\mathrm{Mpc}/h)^3$ and $\log_{10}{P_{\rm shot}}\in [-7,-3]$, respectively.
}
\label{tab:empirical_fit_relations}

\makebox[\textwidth][c]{%
\begin{tabular}{
>{\centering\arraybackslash}m{1.60cm}
*{6}{>{\raggedright\arraybackslash}m{2.85cm}}
}
\toprule
\multirow{2}{*}{Relation}
&
\multicolumn{2}{c}{S40-1024-ref}
&
\multicolumn{2}{c}{S80-1024-ref}
&
\multicolumn{2}{c}{R80-1024-hot}
\\
\cmidrule(lr){2-3}
\cmidrule(lr){4-5}
\cmidrule(lr){6-7}
&
\multicolumn{1}{c}{$P_{\rm shot}$}
&
\multicolumn{1}{c}{$\log_{10}{P_{\rm shot}}$}
&
\multicolumn{1}{c}{$P_{\rm shot}$}
&
\multicolumn{1}{c}{$\log_{10}{P_{\rm shot}}$}
&
\multicolumn{1}{c}{$P_{\rm shot}$}
&
\multicolumn{1}{c}{$\log_{10}{P_{\rm shot}}$}
\\
\midrule

\paramcellfive{$b_\eta(b_\mathrm{F})$}
&
\begin{tabular}[c]{@{}l@{}}
$A=\valerr{0.346}{0.011}{0.009}$\\
$b_\mathrm{F}^{\rm br}=\valerr{0.180}{0.028}{0.025}$\\
$\alpha=\valerr{0.870}{0.114}{0.087}$\\
$\beta=\valerr{0.232}{0.085}{0.093}$\\
$\sigma_{\rm i}<0.0125\ (95\%)$
\end{tabular}
&
\begin{tabular}[c]{@{}l@{}}
$A=\valerr{0.334}{0.011}{0.010}$\\
$b_\mathrm{F}^{\rm br}=\valerr{0.189}{0.023}{0.021}$\\
$\alpha=\valerr{0.843}{0.108}{0.093}$\\
$\beta=\valerr{0.092}{0.075}{0.060}$\\
$\sigma_{\rm i}<0.0150\ (95\%)$
\end{tabular}
&
\begin{tabular}[c]{@{}l@{}}
$A=\valerr{0.301}{0.007}{0.006}$\\
$b_\mathrm{F}^{\rm br}=\valerr{0.231}{0.021}{0.032}$\\
$\alpha=\valerr{0.795}{0.118}{0.077}$\\
$\beta=\valerr{0.138}{0.055}{0.059}$\\
$\sigma_{\rm i}=\valerr{0.0104}{0.0028}{0.0023}$
\end{tabular}
&
\begin{tabular}[c]{@{}l@{}}
$A=\valerr{0.286}{0.007}{0.005}$\\
$b_\mathrm{F}^{\rm br}=\valerr{0.220}{0.033}{0.020}$\\
$\alpha=\valerr{0.807}{0.136}{0.114}$\\
$\beta=\valerr{0.072}{0.042}{0.043}$\\
$\sigma_{\rm i}=\valerr{0.0126}{0.0031}{0.0026}$
\end{tabular}
&
\begin{tabular}[c]{@{}l@{}}
$A=\valerr{0.328}{0.006}{0.007}$\\
$b_\mathrm{F}^{\rm br}=\valerr{0.258}{0.018}{0.032}$\\
$\alpha=\valerr{0.813}{0.084}{0.060}$\\
$\beta=\valerr{0.186}{0.055}{0.051}$\\
$\sigma_{\rm i}=\valerr{0.0095}{0.0028}{0.0023}$
\end{tabular}
&
\begin{tabular}[c]{@{}l@{}}
$A=\valerr{0.314}{0.007}{0.007}$\\
$b_\mathrm{F}^{\rm br}=\valerr{0.266}{0.022}{0.038}$\\
$\alpha=\valerr{0.783}{0.119}{0.076}$\\
$\beta=\valerr{0.106}{0.050}{0.050}$\\
$\sigma_{\rm i}=\valerr{0.0119}{0.0031}{0.0026}$
\end{tabular}
\\

\midrule

\paramcellthree{$c_0^{\alpha}(b_\mathrm{F})$}
&
\begin{tabular}[c]{@{}l@{}}
$A=\valerr{0.233}{0.015}{0.015}$\\
$\gamma=\valerr{1.785}{0.113}{0.109}$\\
$\sigma_{\rm i}<0.0202\ (95\%)$
\end{tabular}
&
\begin{tabular}[c]{@{}l@{}}
$A=\valerr{0.0304}{0.0017}{0.0017}$\\
$\gamma=\valerr{1.109}{0.101}{0.093}$\\
$\sigma_{\rm i}<0.0038\ (95\%)$
\end{tabular}
&
\begin{tabular}[c]{@{}l@{}}
$A=\valerr{0.1419}{0.0053}{0.0055}$\\
$\gamma=\valerr{1.825}{0.061}{0.058}$\\
$\sigma_{\rm i}=\valerr{0.0046}{0.0029}{0.0026}$
\end{tabular}
&
\begin{tabular}[c]{@{}l@{}}
$A=\valerr{0.01652}{0.00082}{0.00090}$\\
$\gamma=\valerr{1.426}{0.075}{0.069}$\\
$\sigma_{\rm i}=\valerr{0.00220}{0.00063}{0.00050}$
\end{tabular}
&
\begin{tabular}[c]{@{}l@{}}
$A=\valerr{0.1563}{0.0057}{0.0059}$\\
$\gamma=\valerr{1.852}{0.061}{0.059}$\\
$\sigma_{\rm i}=\valerr{0.0049}{0.0032}{0.0028}$
\end{tabular}
&
\begin{tabular}[c]{@{}l@{}}
$A=\valerr{0.01720}{0.00091}{0.00103}$\\
$\gamma=\valerr{1.467}{0.082}{0.073}$\\
$\sigma_{\rm i}=\valerr{0.00257}{0.00068}{0.00055}$
\end{tabular}
\\

\midrule

\paramcellthree{$c_2^{\alpha}(b_\mathrm{F})$}
&
\begin{tabular}[c]{@{}l@{}}
$A=\valerr{-0.226}{0.017}{0.017}$\\
$\gamma=\valerr{1.338}{0.130}{0.129}$\\
$\sigma_{\rm i}<0.0236\ (95\%)$
\end{tabular}
&
\begin{tabular}[c]{@{}l@{}}
$A=\valerr{-0.0254}{0.0053}{0.0052}$\\
$\gamma=\valerr{0.764}{0.305}{0.301}$\\
$\sigma_{\rm i}<0.0094\ (95\%)$
\end{tabular}
&
\begin{tabular}[c]{@{}l@{}}
$A=\valerr{-0.1341}{0.0046}{0.0046}$\\
$\gamma=\valerr{1.200}{0.057}{0.056}$\\
$\sigma_{\rm i}<0.0085\ (95\%)$
\end{tabular}
&
\begin{tabular}[c]{@{}l@{}}
$A=\valerr{-0.03435}{0.00281}{0.00264}$\\
$\gamma=\valerr{1.041}{0.120}{0.110}$\\
$\sigma_{\rm i}=\valerr{0.00958}{0.00228}{0.00189}$
\end{tabular}
&
\begin{tabular}[c]{@{}l@{}}
$A=\valerr{-0.1493}{0.0052}{0.0051}$\\
$\gamma=\valerr{1.230}{0.057}{0.057}$\\
$\sigma_{\rm i}<0.0095\ (95\%)$
\end{tabular}
&
\begin{tabular}[c]{@{}l@{}}
$A=\valerr{-0.03703}{0.00330}{0.00306}$\\
$\gamma=\valerr{1.048}{0.130}{0.119}$\\
$\sigma_{\rm i}=\valerr{0.01097}{0.00254}{0.00203}$
\end{tabular}
\\

\bottomrule
\end{tabular}
}
\end{table}

\section{Table of best-fit values for S80-1024-ref at $z=2.4$}

\begin{table}[h]
\centering
\small
\setlength{\tabcolsep}{6pt}
\renewcommand{\arraystretch}{1.3}

\newcommand{\valerr}[3]{#1^{+#2}_{-#3}}

\caption{
Table with 68\% credible intervals for all model parameters of the
Ly$\alpha$ forest auto-power spectrum, \Plya, for the S80-1024-ref simulation,
for the two different priors on the shot-noise term considered in this work, at $z=2.4$. The two columns named $P_\mathrm{shot}$ and $\log_{10}{P_\mathrm{shot}}$ follow the same naming convention as in \autoref{tab:empirical_fit_relations}.
}
\label{tab:bestfit_parameters}

\begin{tabular}{l c c}
\toprule
\multirow{2}{*}{Parameters}
&
\multicolumn{2}{c}{S80-1024-ref}
\\
\cmidrule(lr){2-3}
&
$P_{\rm shot}$
&
$\log_{10}{P_{\rm shot}}$
\\
\midrule

$b_\mathrm{F}$
&
$\valerr{-0.1247}{0.0019}{0.0018}$
&
$\valerr{-0.1279}{0.0017}{0.0017}$
\\

$b_\eta$
&
$\valerr{-0.1868}{0.0037}{0.0037}$
&
$\valerr{-0.1861}{0.0038}{0.0038}$
\\

$c_0^{\alpha}$
&
$\valerr{0.0313}{0.0072}{0.0072}$
&
$\valerr{0.00745}{0.00087}{0.00087}$
\\

$c_2^{\alpha}$
&
$\valerr{-0.0493}{0.0087}{0.0087}$
&
$\valerr{-0.0224}{0.0028}{0.0028}$
\\

$c_4^{\alpha}$
&
$\valerr{0.0161}{0.0048}{0.0047}$
&
$\valerr{0.00190}{0.00165}{0.00165}$
\\

$P_{\rm shot}$ / $\log_{10}{P_{\rm shot}}$
&
$\valerr{-0.0479}{0.0141}{0.0141}$
&
$\valerr{-5.056}{1.342}{1.338}$
\\

$s_0^{\alpha}$
&
$\valerr{0.0061}{0.0044}{0.0044}$
&
$\valerr{-0.0079}{0.0014}{0.0014}$
\\

$s_2^{\alpha}$
&
$\valerr{0.0226}{0.0043}{0.0043}$
&
$\valerr{0.0251}{0.0043}{0.0044}$
\\

\bottomrule
\end{tabular}
\end{table}

\clearpage
\bibliographystyle{utcaps}
\bibliography{Bibliography}
\end{document}